\documentclass[aps,showpacs,preprint,nofootinbib,preprintnumbers]{revtex4-1}

\usepackage{graphicx}
\usepackage{amsmath,amssymb}
\usepackage{amsfonts}
\usepackage{mathrsfs}
\usepackage{bm}
\usepackage{braket}
\usepackage{url}
\usepackage[vcentermath]{youngtab}
\allowdisplaybreaks[3]
\usepackage{hyperref}

\def\arraystretch{1.0}

\newcommand{\be}{\begin{eqnarray}}
\newcommand{\ee}{\end{eqnarray}}

\newcommand{\vslash}{{v\hspace{-5.4pt}/}}

\usepackage[normalem]{ulem}  
\usepackage{color}
\renewcommand\sout{\bgroup \color{red} \ULdepth=-.5ex \ULset}

\begin{document}
\preprint{J-PARC-TH-0106, RIKEN-NC-NP-180, RIKEN-QHP-322}

\title{
Hidden-charm and bottom meson-baryon molecules coupled with five-quark states
}

\author{Yasuhiro Yamaguchi$^{1,2}$}
\author{Alessandro Giachino$^{2,3}$}
\author{Atsushi Hosaka$^{4,5}$}
\author{Elena Santopinto$^{2}$}
\author{Sachiko Takeuchi$^{6,1,4}$}
\author{Makoto Takizawa$^{7,1,8}$}

\affiliation{$^1$
Theoretical Research Division, Nishina Center, RIKEN, Hirosawa, Wako, Saitama 351-0198, Japan
}
\affiliation{$^2$Istituto Nazionale di Fisica Nucleare (INFN), Sezione di
Genova, via Dodecaneso 33, 16146 Genova, Italy}
\affiliation{$^3$Dipartimento di Fisica dell'Universit\`a di Genova, via Dodecaneso 33, 16146 Genova, Italy
}
\affiliation{$^4$Research Center for Nuclear Physics (RCNP), Osaka
University, Ibaraki, Osaka 567-0047, Japan}
\affiliation{$^5$Advanced Science Research Center, Japan Atomic Energy Agency, Tokai, Ibaraki 319-1195, Japan}
\affiliation{$^6$Japan College of Social Work, Kiyose, Tokyo 204-8555, Japan}
\affiliation{$^7$Showa Pharmaceutical University, Machida, Tokyo
194-8543, Japan}
\affiliation{$^8$J-PARC Branch, KEK Theory Center, Institute for Particle and Nuclear Studies, KEK, Tokai, Ibaraki 319-1106, Japan}
\date{\today}

   \begin{abstract}
    In this paper,  we investigate the hidden-charm pentaquarks 
    as $\bar{D}^{(\ast)}\Lambda_{\rm c}$ and
    $\bar{D}^{(\ast)}\Sigma^{(\ast)}_{\rm c}$  molecules coupled to the
    five-quark states.
    Furthermore, we extend our calculations to the hidden-bottom sector.
    The coupling to the five-quark states is treated as the
    short range potential, where
    the relative strength for the meson-baryon channels
    is determined by the structure of the five-quark states.    
    We found that resonant and/or bound states appear 
    in both the charm and bottom sectors. 
    The five-quark state potential turned out to be attractive and, for this reason, it plays an important role to produce these states.
    In the charm sector,  
    we need the five-quark potential in addition to 
    the pion exchange potential
    in producing bound and resonant states, whereas, in the bottom sector, the pion
    exchange interaction is strong enough to produce states.   
    Thus, from this investigation, it emerges that the hidden-bottom pentaquarks are more likely to
    form than their hidden-charm counterparts; 
    for this reason, we suggest that the experimentalists should look for 
    states in the bottom sector.
   \end{abstract}

  \maketitle
  \tableofcontents

  \section{Introduction}
  The study of the exotic hadrons has aroused great interest in 
  nuclear and hadron physics.
  In 2015, the Large Hadron Collider beauty experiment (LHCb) collaboration observed two hidden-charm
  pentaquarks, $P^+_c(4380)$ and $P^+_c(4450)$, in
  $\Lambda^0_b\rightarrow J/\psi K^-p$ decay~\cite{Aaij:2015tga,Aaij:2016phn,Aaij:2016ymb}.
  These two pentaquark states are found to have masses of
  $4380\pm8\pm28\;$ MeV and $4449.8\pm 1.7 \pm 2.5\; $ MeV,
  with corresponding widths of
  $205\pm18\pm86 \; $ MeV and $39\pm5\pm19\; $ MeV. 
  The spin-parity $J^P$ of these states has not yet been determined.  
  The parities of these states are preferred to be opposite, and one state has $J=3/2$ and the other $J=5/2$.
  $(J^P_{P^+_c(4380)},J^P_{P^+_c(4450)})=(3/2^-,5/2^+)$ gives the best fit solution, but $(3/2^+,5/2^-)$
  and $(5/2^-,3/2^+)$ are also acceptable.
  The $P^{+}_{\rm c}$ resonances are one of topics of great interest 
  as the candidates of the exotic multiquark state,
  and 
  many discussions
  have been done so far~\cite{Chen:2016qju,Ali:2017jda,Esposito:2016noz}.

   Hidden-charm pentaquark states, such as $uudc\bar{c}$ and $udsc\bar{c}$
   compact structures, have been studied so far.
   Before $P^+_{\rm c}$ observed by LHCb, 
   Yuan {\it et al.} in \cite{Yuan:2012wz}
   studied the $uudc\bar{c}$ and $udsc\bar{c}$ systems by the
   non-relativistic harmonic oscillator Hamiltonian with three kinds of
   the schematic interactions:
   a chromomagnetic interaction, a flavor-spin-dependent interaction and an instanton-induced interaction.
   In \cite{Santopinto:2016pkp}, 
   Santopinto {\it et al.} investigated the hidden-charm pentaquark states as five-quark compact states in the $S-$wave
   by using a constituent quark model approach.  
   The hidden-charm and hidden-bottom pentaquark masses  have been calculated by 
   Wu {\it et al.} in \cite{Wu:2017weo}, by means of
   a color-magnetic interaction between the three light quarks and the
   $c\bar{c}$ ($b\bar{b}$) pair in a color octet state. %
   Takeuchi {\it et al.}~\cite{Takeuchi:2016ejt} has also investigated the hidden-charm pentaquark states 
   by the quark cluster model, and discussed the structure of the five-quark states 
   which appears in
   the scattering states.
   To investigate the compact five-quark state,
   the diquark model has also been applied~\cite{Maiani:2015vwa,Li:2015gta,
   Ali:2016dkf,Lebed:2015tna,Zhu:2015bba,Zhu:2015bba}.
   The quantum chromodynamics (QCD) sum rules with the diquark picture were applied in Refs.~\cite{Wang:2015epa,Wang:2015ava}.
   However, these authors do not provide any information about the pentaquark widths.
   Despite many theoretical works and implications, there is so far no
   clear evidence of such compact multiquark states.

  By contrast, it is widely accepted that there are  candidates for hadronic molecular  states.  
  A long-standing and well-known example is 
  $\Lambda(1405)$, which is considered to be a molecule of 
  $\bar{K}N$ and $\pi\Sigma$ coupled channels.  
  A general review of $\Lambda(1405)$ can be found in~\cite{Hyodo:2011ur} .
  In the heavy quark sector, 
  $X(3872)$~\cite{Choi:2003ue},   
  $Z_b(10610)$,  and  $Z_b(10650)$~\cite{Collaboration:2011gja}
   are considered to be, respectively,
  $D\bar{D}^\ast$~\cite{Tornqvist:2004qy,Close:2003sg,Braaten:2003he,Wong:2003xk,Swanson:2003tb,Swanson:2004pp}
  and $B^{(\ast)}\bar{B}^\ast$ molecules~\cite{Bondar:2011ev,Ohkoda:2011vj}.
  Now, the $P^+_c$ pentaquarks have been found just below the $\bar{D}\Sigma^\ast_{\rm c}$
  and $\bar{D}^\ast\Sigma_{\rm c}$ thresholds. Thus, the $\bar{D}\Sigma^\ast_{\rm c}$ and $\bar{D}^\ast\Sigma_{\rm
  c}$ molecular components are expected to be
  dominant~\cite{Wu:2010jy,Wu:2010vk,Garcia-Recio:2013gaa,Karliner:2015ina,Chen:2015loa,Roca:2015dva,He:2015cea,Meissner:2015mza,Chen:2015moa,Uchino:2015uha,Burns:2015dwa,Shimizu:2016rrd,Yamaguchi:2016ote,Shimizu:2017xrg}.  
  Moreover, the baryocharmonium structure as the composite of $J/\psi$ and
  the excited nucleon $N^\ast$ is also discussed~\cite{Kubarovsky:2015aaa}.  
  
  In the formation of the hadronic molecules, 
  the one pion exchange potential (OPEP) would be a key ingredient to bind the composite hadrons. 
  In nuclear physics, 
  it has been well-known that the pion-exchange is a driving force to
  bind atomic nuclei~\cite{Ikeda:2010aq}.
  Moreover, 
  it
  was also applied to
  the deuteronlike bound states of two hadrons, which is called deusons~\cite{Tornqvist:1993ng}.
  Specifically in the heavy quark sector, the role of the pion-exchange
  would be enhanced by the heavy quark spin symmetry.   
  The important property of this symmetry is that in the heavy quark
  mass limit, the spin of heavy (anti)quarks, $s_Q$, is decoupled from  
  the total angular momentum of the light degrees of freedom, $j$,
  which is carried by light quarks and gluons~\cite{Isgur:1989vq,Isgur:1989ed,Isgur:1991wq,Neubert:1993mb,Manohar:2000dt,Yasui:2013vca,Yamaguchi:2014era,Hosaka:2016ypm}.
  Thus, the heavy quark spin (HQS) multiplet emerges, 
  where hadrons in the multiplet have the same mass,
  even though the hadrons have different total angular momenta given by $s_Q\otimes j$.
  In the charm (bottom) mesons, 
  a $\bar{D}$ ($B$) meson\footnote{Actually, $\bar{D}$ ($B$) is the anti-charm (anti-bottom) meson including anti-charm (anti-bottom) quark with charm (bottom) number $= -1$. In this paper, however, we just call them the charm (bottom) meson.} as a pseudoscalar meson is regarded as the member of the HQS doublet whose pair is a $\bar{D}^\ast$ ($B^\ast$) meson as a vector meson.  
  In fact, the mass difference of $\bar{D}$ and $\bar{D}^\ast$ mesons ($B$ and $B^\ast$ mesons)
  is small, $m_{\bar{D}^\ast}-m_{\bar{D}}\sim 140$ MeV ($m_{B^\ast}-m_{B}\sim 45$ MeV).
  In contrast, the mass differences in the light flavor sectors
  are given by $m_\rho-m_\pi\sim 630$ MeV and $m_{K^\ast}-m_{K}\sim 390$ MeV.
  The approximate mass degeneracy 
  enhances the attraction due to
  the mixing of the $\bar{D}$ ($B$) meson and the
  $\bar{D}^\ast$ ($B^\ast$) meson 
  caused by the pion-exchange.
  We note that the heavy meson is coupled to the pion through
  the $\bar{D}^\ast\bar{D}\pi$ and $\bar{D}^\ast\bar{D}^\ast\pi$
  couplings, while the $\bar{D}\bar{D}\pi$ coupling is absent due to
  the parity and angular momentum conservation.
  In the systems of the heavy meson and nucleon,
  the attraction of the pion-exchange via the process
  $\bar{D}N\leftrightarrow\bar{D}^\ast N$ ($BN\leftrightarrow B^\ast N$)
  was discussed (See review in Ref.~\cite{Hosaka:2016ypm} and references
  therein).

  Similarly, in the heavy-light baryons,
  $\Sigma_{\rm c}$ ($\Sigma_{\rm b}$) and $\Sigma^\ast_{\rm c}$ ($\Sigma^\ast_{\rm b}$)
  belong to the HQS doublet, where the mass difference of the baryons is given by 
  $m_{\Sigma^\ast_{\rm c}}-m_{\Sigma_{\rm c}}\sim 65$ MeV
  ($m_{\Sigma^\ast_{\rm b}}-m_{\Sigma_{\rm b}}\sim 20$ MeV).
  On the other hand, a $\Lambda_{\rm c}$ ($\Lambda_{\rm b}$) baryon
  belongs to the HQS singlet,
  because the spin of the light diquark is zero. 
  The heavy quark spin symmetry yields that the thresholds of $\bar{D}\Sigma_{\rm c}$,
  $\bar{D}\Sigma^\ast_{\rm c}$, $\bar{D}^\ast\Sigma_{\rm c}$, and
  $\bar{D}^\ast\Sigma^\ast_{\rm c}$ are close to each other.
  In addition, the $\bar{D}\Lambda_{\rm c}$ and
  $\bar{D}^\ast\Lambda_{\rm c}$ thresholds are also located just below the
  $\bar{D}^{(\ast)}\Sigma^{(\ast)}_{\rm c}$.
  Thus, the meson-baryon system should be a coupled-channel system,
  and the spin-dependent operator of the pion-exchange potential has a
  role to mix the above various channels.

  Among these molecular candidates, the most explored $X(3872)$ is also known to be produced 
by high-energy $p\bar{p}$ collisions \cite{Acosta:2003zx,Abazov:2004kp}, which implies an admixture of a compact 
and a molecular component~\cite{Hosaka:2016pey}.  
  The admixture structure of hadrons is eventually a rather conceptual problem 
of compositeness of hadrons as discussed long ago in~\cite{Weinberg:1965zz,Weinberg:1962hj,Lurie:1964ab}
and recently in~\cite{Nagahiro:2013hba,Nagahiro:2014mba,Yamaguchi:2016kxa,Sekihara:2014kya,Kamiya:2016oao}.
However, it provides a useful framework 
to solve efficiently 
complicated problems
when using
quarks and gluons of QCD directly.
Indeed, the nontrivial properties of $X(3872)$ may be explained 
by this
admixture picture of a $c\bar{c}$ core plus higher Fock components due to the coupling to the meson-meson continuum~\cite{Hosaka:2016pey,Kalashnikova:2005ui,Suzuki:2005ha,Barnes:2007xu,Zhang:2009bv,Matheus:2009vq,Kalashnikova:2009gt,Ortega:2010qq,Danilkin:2010cc,Coito:2010if,Coito:2012vf,Ferretti:2013faa,Chen:2013pya,Takizawa:2012hy,Takeuchi:2014rsa}.
  For those interested in $X$, $Y$, and $Z$ exotic states, a general review can be found in~\cite{Hosaka:2016pey}. 
  In general, if more than one 
  state is   
  allowed for a given set of quantum numbers,
  the hadronic resonant states are unavoidably mixtures of these
  states. Therefore, an important issue is to clarify how these components are mixed in 
  physical hadrons.


One of the best approaches to gaining insight into the nature of the pentaquark states consists 
of producing these  states in a different reaction. 
In particular,  the case of prompt production is  important because a positive answer 
will 
indicate that the pentaquark has a compact
nature,  while 
a negative answer will not 
exclude
the
pentaquark as a molecular state.
For example, a particular kind of prompt production is 
photoproduction, which was first proposed by Wang in \cite{Wang:2015jsa} 
to investigate the nature of the pentaquark states.
A search for LHCb-pentaquark will be carried out at Jefferson Lab in exclusive $J/\psi$ production off protons 
by real (Hall A/C)~\cite{Meziani:2016lhg} and quasi-real (Hall-B)~\cite{J-LAB_E12-12-001,J-LAB_Hall_B_proposals} photons.
Moreover, two electroproduction experiments have been proposed in
the same facility.
Prompt production experiments may also be proposed at
CERN, KEK, GSI-FAIR, and J-PARC.
There have also been theoretical discussions about the pentaquark productions
via the photoproduction~\cite{Huang:2013mua,Huang:2016tcr},
the pion-nucleon collision~\cite{Garzon:2015zva,Liu:2016dli,Kim:2016cxr},
and 
the $p\bar{p}$ collision~\cite{Wu:2010jy,Wu:2010vk}.
The studies from both experimental and theoretical sides are also
important to know that the LHCb data shows whether a resonance structure or
a kinematic effect as discussed in Refs.~\cite{Guo:2015umn,Liu:2015fea,Mikhasenko:2015vca}.

Those discussions of the hidden-charm pentaquarks can be extended to those of the hidden-bottom partners.
The hidden-bottom partner would be easy to 
be formed,
because 
the kinetic term should be suppressed due to the large hadron masses.
Moreover, we expect that the small mass splittings of $B$ and $B^\ast$, and $\Sigma_{\rm b}$ and $\Sigma^\ast_{\rm b}$
induce the strong coupled channel effect.
   The mass and production of the hidden-bottom pentaquarks have been studied 
   in Refs.~\cite{Chen:2016qju,Wu:2017weo,Shimizu:2016rrd,Wu:2010rv,Xiao:2013jla,Azizi:2017bgs,Cheng:2016ddp}.

  In this paper,  we investigate the  hidden-charm pentaquarks   as  $\bar{D}^{(\ast)}\Lambda_{\rm c}$ and
   $\bar{D}^{(\ast)}\Sigma^{(\ast)}_{\rm c}$  molecules coupled to the
  five-quark states. 
  The inclusion of the five-quark state is inspired by the recent work 
  of 
  Takeuchi {\it et al.}~\cite{Takeuchi:2016ejt}  
  by means of the quark cluster model.  
   Moreover, we extend our calculations 
   to the hidden-bottom
   sector. 
   We provide predictions  for  hidden-bottom pentaquarks 
    as ${B}^{(\ast)}\Lambda_{\rm b}$ and
  ${B}^{(\ast)}\Sigma^{(\ast)}_{\rm b}$  molecules coupled to the
  five-quark states.  
     Here, $\bar{D}^{(\ast)}$ ($\Sigma^{(\ast)}_{\rm c}$) stands for $\bar{D}$ and $\bar{D}^\ast$
   ($\Sigma_{\rm c}$ and $\Sigma^{\ast}_{\rm c}$), while  ${B}^{(\ast)}$ ($\Sigma^{(\ast)}_{\rm b}$) stands for ${B}$ and ${B}^\ast$
   ($\Sigma_{\rm b}$ and $\Sigma^{\ast}_{\rm b}$).
   Coupling to the five-quark states is described as the short-range
   potential between the meson and the baryon.    We also introduce the long-range force given by the one-pion exchange potential.
     By solving the coupled channel Schr\"odinger equation, we study the bound and resonant hidden-charm and hidden-bottom pentaquark states for $J^P=\frac{1}{2}^-$,
   $\frac{3}{2}^-$, and $\frac{5}{2}^-$ with isospin $I=\frac{1}{2}$.   \\

  This paper is organized as follows.
  In Section~\ref{sec:Modelsetup}, 
  we introduce our coupled-channel model.
  Specifically,  in Section~\ref{sec:MBand5q}, 
  the meson-baryon and the five-quark channels are introduced, while in
  Sections~\ref{sec:OPEP} 
  and \ref{sec:Couplings_to_5q}, 
  respectively,  the OPEP as the long-range force,
  and the five-quark state as the  short-range force are presented.  
 The model parameters, the numerical methods, and the results for the hidden-charm and 
the hidden-bottom sectors are discussed  in Sections
 \ref{sec:Model_parameters}, 
 \ref{sec:Numerical_methods}, 
 \ref{sec:Numerical_results_charm}, 
 and \ref{sec:Numerical_results_bottom}, 
 respectively, while in Section~\ref{sec:Comparison_QCM}, 
 we compare, for the hidden-charm sector, our numerical results with 
 those of 
 the quark cluster model by Takeuchi~\cite{Takeuchi:2016ejt},
 and find that they are similar to each other.
 In Section~\ref{sec:Numerical_results_bottom}, 
 we discuss the idea 
that in the hidden-bottom sector,
 we expect to provide reliable predictions
 for the hidden-bottom pentaquark masses and widths, which will be useful for future experiments.
 We also discuss 
that the hidden-bottom pentaquarks  are  more  likely  to  form than their hidden-charm counterparts; 
for this reason, we suggest that the experimentalists should look for these states.
Finally, Section~\ref{sec:summary} summarizes the work as a whole.

  \section{Model setup}
  \label{sec:Modelsetup}

  \subsection{Meson-baryon and $5q$ channels}
  \label{sec:MBand5q}
So far many studies for exotic states have been performed by using various models such as hadronic molecules,
compact multi-quark states, hybrids with gluons and so on.  
Strictly in QCD, definitions of these model states are not trivial, while the physical exotic states appear  
as resonances in scatterings of hadrons. 
Therefore, the issue is 
related to the question of the compositeness  of resonances,
which has been discussed for a long time~\cite{Weinberg:1965zz,Weinberg:1962hj,Lurie:1964ab}, 
and recently in the context of hadron resonances
(see for instance~\cite{Nagahiro:2013hba,Nagahiro:2014mba} and references therein).
In nuclear physics a similar issue has been discussed in the context of 
clustering phenomena of 
nuclei~\cite{Funaki:2009zz}.
In the end, it 
comes down 
to the question of efficiency in solving the complex many-body systems.  
In the current problem of pentaquark $P_c$, there are two competing sets of channels:
the meson-baryon ($MB$) channels and the five-quark ($5q$) channels\footnote{Various combinations of hadrons and quark configurations which may form the pentaquark $P_c$ are called channels. }.  

The meson-baryon channels describe the dynamics at long distances.  
The base states  may be formed by open-charm hadrons, such as
$\bar{D}^*\Sigma_c$, and hidden ones, such as $J/\psi N$.  
Considering the mass of the observed $P_c$, which is much closer to the open-charm channels
than to the hidden ones, we may neglect the hidden-charm channels at the
first attempt.
However, the 
hidden-charm channels 
become important when discussing decays 
of possible pentaquark states, such as 
the $J/\psi N$ observed in the LHCb experiment.
For the hidden-bottom sector, however, the thresholds
between the open-bottom meson-baryon channel and the $\Upsilon(1S)N$ are
rather different, the order of 500 MeV.
Therefore, the $\Upsilon(1S)N$ component seems to be suppressed in the
hidden-bottom pentaquarks.
On the other hand, the threshold of $\Upsilon(2S)N$ 
is close to the open-bottom thresholds.
Experimentally, the measurement in the open-bottom meson-baryon and $\Upsilon(2S)N$ decays is
preferred rather than that in the $\Upsilon(1S)N$ decay. 
Our model space for open charm hadrons are summarized in Table~\ref{table:MBchannels}.  
For the interaction between them, we employ 
the one-pion exchange potential, which is  
the best established interaction due to chiral symmetry and its spontaneous breaking.  
Explicit forms of the potential are given in Appendix~\ref{sec:appendix_OPEP}. 

The $5q$ part describes the dynamics at short distances, which we consider 
to be in 
the order of 1 fm or less.  
Inspired by the recent discussion~\cite{Takeuchi:2016ejt}, 
we consider $5q$ compact states formed 
by color-octet light quarks 
($3q$) and color octet $c \bar c$.  
The relevant channels are summarized in Table~\ref{table:5qchannels}.
Notations are $[q^3 D_C, S_{3q}]S_{c\bar c}$
where $D_C = 8$ indicates that $qqq$ form the color octet,
$S_{3q}$
is the spin of the light quarks $qqq=uud$, and 
$S_{c\bar c}$  the spin of $c \bar c$.
This $5q$ channel is considered to be the lowest eigenstate, for example, 
of the breathing mode of the five-quarks, 
which has the overlap with the meson-baryon channel but should be included separately in the system.

\begin{table}[htbp]
 \caption{Various channels of open-charm meson-baryons of total spin parity $J^P$ with
 $^{2S+1}L$.
 }
 \label{table:MBchannels}
 \begin{center}
  \begin{tabular}{c  l l l l l l }
   \hline\hline
   ${\rm Channels}$  
       & $\bar{D}\Lambda_{\rm c}$
       & $\bar{D}^\ast\Lambda_{\rm c}$
       & $\bar{D}\Sigma_{\rm c}$
       & $\bar{D}\Sigma^\ast_{\rm c}$
       & $\bar{D}^\ast\Sigma_{\rm c}$
       & $\bar{D}^\ast\Sigma^\ast_{\rm c}$\\
  \hline
$J^P$  & & & & & & \\ 
$1/2^-$
       & $^2S$
       & $^2S,\ ^{4 \hspace{-0mm}}D$
       & $^2S$
       & $^{4 \hspace{-0mm}}D$
       & $^2S,\ ^{4 \hspace{-0mm}}D$
       & $^2S,\ ^{4 \hspace{-0mm}}D,\ ^{6 \hspace*{-0mm}}D$ \\
$3/2^-$
       & $^2D $
       & $^4S,\ ^2D,\ ^{4 \hspace{-0mm}}D$
       & $^2D$
       & $^4S,\ ^{4 \hspace{-0mm}}D$
       & $^4S,\ ^2D,\ ^{4 \hspace{-0mm}}D$
       & $^4S,\ ^2D,\ ^{4 \hspace{-0mm}}D,\ ^{6 \hspace*{-0mm}}D,\ ^{6 \hspace*{-0mm}}G$ \\
$5/2^-$
       & $^2D$
       & $^2D,\ ^4D,\ ^4G$
       & $^2D$
       & $^{4 \hspace{-0mm}}D,\ ^4G$
       & $^2D,\ ^{4 \hspace{-0mm}}D,^4G$
       & $^6S,\ ^2D,\ ^{4 \hspace{-0mm}}D,  ^6D,\ ^4G,\ ^{6 \hspace*{-0mm}}G$ \\
   \hline\hline
  \end{tabular}
 \end{center}
\end{table}

\renewcommand\arraystretch{1.5}
\begin{table}[htbp]
 \caption{Channels of 5$q$'s with color octet $qqq$ and $c\bar c$
with possible total spin $J$.  For notations, see text. 
 }
 \label{table:5qchannels}
 \begin{center}
  \begin{tabular}{c  l l l l }
   \hline\hline
   Channel
       & $[ q^3 8, \frac{1}{2}]0$ 
       & $ [ q^3 8, \frac{1}{2}]1$ 
       & $ [ q^3 8, \frac{3}{2}]0$ 
       & $ [ q^3 8, \frac{3}{2}]1$ 
\\
  \hline
 $J$
       & 1/2
       & 1/2, 3/2
       & 3/2
       & 1/2, 3/2, 5/2
\\ 
   \hline\hline
  \end{tabular}
 \end{center}
\end{table}

Thus, our model Hamiltonian, expanded by the open-charm $MB$  and $5q$ channels, is written as
\be
H = 
\left(
\begin{array}{c c }
H^{MB} & V \\
V^\dagger & H^{5q}
\end{array}
\right)
\label{eq:Ham_MB-5q}
\ee
where the $MB$ part $H^{MB}$ contains $K_i$; the kinetic energy of each $MB$ channel $i$ and 
$V^{\pi}_{ij}$; the OPEP potential, 
and $H^{5q}$ 
stands 
for the $5q$ channels.  
For simplicity, we consider that $H^{5q}$ is diagonalized by the 
$5q$ channels (denoted by $\alpha$) of Table~\ref{table:5qchannels}
and its eigenvalue is expressed by $M_{\alpha}$. 
The off-diagonal part in~\eqref{eq:Ham_MB-5q}, $V$,
represents 
the transition between the 
$MB$ and $5q$ channels.  
In the quark cluster model, such interactions are modeled 
by quark exchanges accompanied by gluon exchanges.  
In the present paper, we shall make a simple assumption
that ratios of transitions between various channels  $i \sim MB$ and 
$\alpha \sim 5q$ are dominated by the spectroscopic factors, 
overlaps 
$\braket{i | \alpha}$.
The absolute strengths are then assumed to be determined by a single parameter.  
Various components of the Hamiltonian are then written as
\be
\left( H^{MB}_{ij} \right) =
\left(
\begin{array}{c c c}
K_{1} + V_{11}^\pi & V_{12}^{\pi} &  \cdots \\
V_{21}^{\pi} & K_{2} +V_{22}^\pi & \cdots \\
\cdots & \cdots & \cdots
\end{array}
\right)
\, , \; \; \; 
(H^{5q}_{\alpha \beta}) =
\left(
\begin{array}{c c c}
M_1 & 0 &  \cdots \\
0 & M_2 & \cdots \\
\cdots & \cdots & \cdots
\end{array}
\right)
\ee
and
\be
(V_{i \alpha}) =  
(\braket{i | \alpha}) 
=
\left(
\begin{array}{c c c}
V_{11} & V_{12} &  \cdots \\
V_{21} & V_{22} & \cdots \\
\cdots & \cdots & \cdots
\end{array}
\right).
\ee

\begin{figure}[h]
\begin{center}
 \includegraphics[width=0.8 \linewidth]{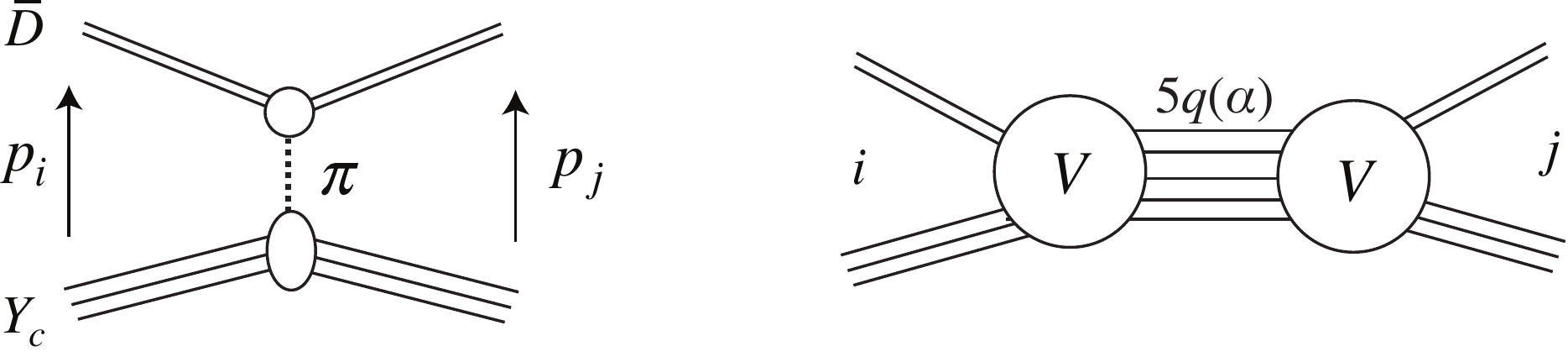}
\end{center}
\vspace{-5mm}
\caption{One pion exchange potential (left) and the effective interaction due to the coupling to the 5$q$ channel (right).   
The meson-baryon channels are generally represented by $\bar D$ and $Y_c$, respectively, and $i$ is for the initial and $j$ the final channels.  A $5q$ channel is denoted by $\alpha$.    
}
\label{fig_OPEP_5q}
\end{figure}

Now let us consider the coupled equation for the $MB$ and $5q$ channels, 
$H \psi = E \psi$, where $\psi = (\psi^{MB}, \psi^{5q})$, 
\be
H^{MB} \psi^{MB} + V \psi^{5q} &=& E \psi^{MB},
\nonumber \\
V^\dagger \psi^{MB} + H^{5q} \psi^{5q} &=& E \psi^{5q}.
\nonumber 
\ee
Solving the second equation for $\psi^{5q}$, 
$
\psi^{5q} = {(E-H^{5q})^{-1}}V^\dagger \psi^{MB}
$
and substituting for the first equation, we find the equation for $\psi^{MB}$, 
\be
\left( K^{MB} + 
V^\pi
+ V \frac{1}{E-H^{5q}}V^\dagger \right)
\psi^{MB} = E \psi^{MB}.
\ee
The last term on the left-hand side is due to the elimination of the $5q$ channels, and is regarded 
as an effective interaction for the $MB$ channels.
Thus, the total interaction for the $MB$ channels is defined by
\be
U = 
V^\pi + V \frac{1}{E-H^{5q}}V^\dagger .
\label{eq_def_Ueff}
\ee
We then insert 
the assumed $5q$ eigenstates into the second term of (\ref{eq_def_Ueff}), 
\be
U_{ij} = 
V^\pi_{ij}
+ \sum_{\alpha}
\Braket{i | V | \alpha} 
\frac{1}{E-E^{5q}_{\alpha}} 
\Braket{\alpha | V^\dagger | j} 
\label{eq:effective_short_range}
\ee
where $E^{5q}_{\alpha}$ is the eigenenergy of a $5q$ channel.  
In this equation, we have indicated the meson-baryon channel by $i,j$,
and $5q$ channels by $\alpha$.
In this way, the effects of the $5q$ channels are included in the form of 
effective short range  interaction.
The corresponding diagram of this equation is shown in Fig.~\ref{fig_OPEP_5q}.  
The computations for the OPEP and the short range interactions are discussed 
in the next sections.  

  \subsection{One pion exchange potential}
  \label{sec:OPEP}
In this subsection, we derive the one pion exchange potential (OPEP)
between $\bar{D}^{(\ast)}$ and $Y_{\rm c}$ in the first term of Eq.~\eqref{eq:effective_short_range}.
Hereafter, we use the notation $\bar{D}^{(\ast)}$ to stand for a
$\bar{D}$ meson, or a $\bar{D}^\ast$ meson, and
$Y_c$ to stand for $\Lambda_{\rm c}$, $\Sigma_{\rm c}$, or $\Sigma^{\ast}_{\rm c}$.

The OPEP is obtained by the effective Lagrangians for heavy mesons
(baryons) and the Nambu-Goldstone boson, satisfying the heavy quark and
chiral symmetries.
The Lagrangians for heavy mesons and the Nambu-Goldstone bosons are
given by~\cite{Wise:1992hn,Burdman:1992gh,Yan:1992gz,Falk:1992cx,Casalbuoni:1996pg,Manohar:2000dt}
\begin{align}
 &{\cal L}_{\pi HH}=g_\pi {\rm Tr}\left[H_b \gamma_\mu\gamma_5 A^{\mu}_{ba}\bar{H}_{a}\right].
 \label{eq:LagrangianDDpi}
\end{align}
The trace ${\rm Tr}\left[\cdots\right]$ is taken over the gamma matrix.
The heavy meson fields $H$ and $\bar{H}$
are represented by
\begin{align}
&
 H_a=\frac{1+\vslash}{2}
 \left[\bar{D}^\ast_{a\mu}\gamma^\mu-\bar{D}_a\gamma_5\right], \\
 &\bar{H}_a=\gamma_0 H^\dagger_a\gamma_0,
\end{align}
where the fields are constructed by the heavy pseudoscalar meson 
$\bar{D}$
and
the vector meson 
$\bar{D}^\ast$
belonging to the heavy quark spin (HQS) doublet.
$v_\mu$ is a four-velocity of a heavy quark, and satisfies $v^\mu v_\mu=1$
and $v^0 >0$.
The subscripts $a,b$ are for the light flavor $u,d$.
The axial vector current for the pion, $A_\mu$, is given by
\begin{align}
 &A_\mu=\frac{i}{2}\left[\xi^\dagger(\partial_\mu
 \xi)+(\partial_\mu \xi)\xi^\dagger\right],
\end{align}
where $\xi=\exp\left(\frac{i\hat{\pi}}{2f_\pi}\right)$ with the pion
decay constant $f_\pi=92.3$ MeV.
The pion field $\hat{\pi}$ is given by
\begin{align}
 \hat{\pi}=
 \sqrt{2}\left(
  \begin{array}{cc}
   \frac{\pi^0}{\sqrt{2}}&\pi^+ \\
   \pi^- & -\frac{\pi^0}{\sqrt{2}} \\
  \end{array}
 \right).
\end{align}
The coupling constant $g_\pi$ is determined by the strong decay of
$D^\ast\rightarrow D\pi$ as $g_\pi=0.59$~\cite{Casalbuoni:1996pg,Manohar:2000dt,Olive:2016xmw}.

The Lagrangians for heavy baryons and Nambu-Goldstone bosons are given
by~\cite{Yan:1992gz,Liu:2011xc}
\begin{align}
 {\cal L}_{\pi
 BB}&=\frac{3}{2}g_1(iv_\kappa)\varepsilon^{\mu\nu\lambda\kappa}{\rm tr}\left[
 \bar{S_\mu}A_\nu S_\lambda\right]
 + g_4{\rm tr}\left[\bar{S}^\mu A_{\mu} B_{\bar{3}}\right] + {\rm H.c.}
 \label{eq:LagrangianBBpi}
\end{align}
The trace ${\rm tr}\left[\cdots\right]$ is for the flavor space.
The superfields $S_\mu$ and $\bar{S}_\mu$
are represented by
\begin{align}
 &S_{\mu}=
\hat{\Sigma}^{\ast}_{{\rm c}\mu}
+\frac{\delta}{\sqrt{3}}\left(\gamma_\mu+v_\mu\right)\gamma_5
 \hat{\Sigma}_{\rm c}, \\
 &\bar{S}_\mu = \gamma_0S^\dagger_\mu \gamma_0,
\end{align}
with the 
$\hat{\Sigma}_{\rm c}$ 
and 
$\hat{\Sigma}^\ast_{\rm c}$ 
fields in the HQS multiplet.
The phase factor $\delta$ is 
set at 
$\delta=-1$, as discussed in Ref.~\cite{Liu:2011xc}.
The heavy baryon fields 
$\hat{\Lambda}_{\rm c}$ and $\hat{\Sigma}^{(\ast)}_{\rm c (\mu)}$ 
are expressed
by
\begin{align}
 &
 \hat{\Lambda}_{\rm c}
 =
\left(
 \begin{array}{cc}
  0&\Lambda^+_{\rm c}  \\
  -\Lambda^+_{\rm c} & 0  \\
 \end{array}
 \right), \quad
 \hat{\Sigma}^{(\ast)}_{\rm c (\mu)}
 =\left(
 \begin{array}{cc}
  \Sigma^{(\ast)++}_{\rm c (\mu)}&\frac{1}{\sqrt{2}}\Sigma^{(\ast)+}_{\rm c (\mu)}  \\
  \frac{1}{\sqrt{2}}\Sigma^{(\ast)+}_{\rm c (\mu)} &\Sigma^{(\ast)0}_{\rm c (\mu)}  \\
 \end{array}
 \right).
\end{align}
The coupling constants $g_1$ and $g_4$, given as
$g_1=(\sqrt{8}/3)g_4=1$, are used, 
which are obtained by the quark model estimation discussed in Ref.~\cite{Liu:2011xc}.
For the coupling $g_4$, this value can also be fixed by the $\Sigma^{(\ast)}_{\rm c} \to \Lambda_{\rm c} \pi$ decay,
and agrees with the one obtained by the quark model~\cite{Liu:2011xc}.

For the hidden-bottom sector, 
these effective Lagrangians are also applied by replacing the charmed hadron fields 
by
the bottom hadron fields, while the same coupling constants are used.

In order to parametrize the internal structure of hadrons, we introduce
the dipole form factor at each vertex:
\begin{align}
 &F(\Lambda,\vec{q}\,)=
 \frac{\Lambda^2-m^2_\pi}{\Lambda^2+\vec{q}\,^2},
 \label{eq:formfactor}
\end{align}
with the pion mass $m_\pi$ and the three-momentum $\vec{q}$ of an
incoming pion.
As discussed in
Refs.~\cite{Yasui:2009bz,Yamaguchi:2011xb,Yamaguchi:2011qw},
the cutoffs of heavy hadrons are fixed by the ratio between the sizes of
the heavy hadron and nucleon, $\Lambda_N/\Lambda_{H}=r_{H}/r_N$
with the cutoff and size of the heavy hadron being $\Lambda_H$ and
$r_H$, respectively.
The nucleon cutoff is determined to reproduce the deuteron-binding
energy by the OPEP
as $\Lambda_N=837$ MeV~\cite{Yasui:2009bz,Yamaguchi:2011xb,Yamaguchi:2011qw}.
The ratios are computed by the means of constituent quark model with
the harmonic oscillator potential~\cite{Oh:2009zj}, where
the frequency is evaluated by the hadron charge radii in Refs.~\cite{Hwang:2001th,SilvestreBrac:1996bg}.
For the heavy meson~\cite{Yasui:2009bz}, 
we obtain 
$\Lambda_{\bar{D}}=1.35\Lambda_N$ and
$\Lambda_{B}=1.29\Lambda_N$ for the $\bar{D}^{(\ast)}$ meson and the $B$
meson, respectively.
For the heavy baryon~\cite{Oh:2009zj}, we obtain 
$\Lambda_{\Lambda_{\rm c}} \sim \Lambda_{\Sigma_{\rm c}} \sim \Lambda_N$ for the charmed baryon,
and $\Lambda_{\Lambda_{\rm b}} \sim \Lambda_{\Sigma_{\rm b}} \sim \Lambda_N$ 
for the bottom baryon.
We note that values of these cutoffs are smaller than those used in other
studies, e.g. $\Lambda=2.35$ GeV and $\Lambda=1.77$ GeV in Ref.~\cite{Chen:2015loa}.

From these Lagrangians \eqref{eq:LagrangianDDpi} and
\eqref{eq:LagrangianBBpi}, and the form factor~\eqref{eq:formfactor},
we obtain the OPEP as the Born term of the
scattering amplitude.
The explicit form of the OPEP is summarized in Appendix~\ref{sec:appendix_OPEP}.
The OPEP is also used for the hidden-bottom sector, $B^{(\ast)}Y_b$,
by employing the cutoff parameters $\Lambda_B$, $\Lambda_{\Lambda_{\rm b}}$,
and $\Lambda_{\Sigma_{\rm b}}$, where $B^{(\ast)}$ stands for $B$
or $B^\ast$, and $Y_b$ stands for $\Lambda_{\rm b}$,
$\Sigma_{\rm b}$ or $\Sigma^\ast_{\rm b}$.
Let us remark about the contact term of the OPEP. In this study, it is neglected as shown in Eq.~\eqref{eq:central_force}
as is in the conventional nuclear physics. 
We assume that the OPEP appears only in the long range hadronic region. 
As discussed above, the cutoff parameters of the OPEP are determined from 
the ratio of sizes of the relevant hadron and nucleon.
The cutoff of the nucleon is determined so as to reproduce the deuteron binding energy without the contact term~\cite{Yasui:2009bz}.
  
  \subsection{Couplings to $5q$ states}
  \label{sec:Couplings_to_5q}
In this subsection, we derive 
the effective short-range interaction, the 2nd term of 
 (\ref{eq:effective_short_range}).
To do so,  we need to know the matrix elements 
   $\Braket{ i | V | \alpha } $
   and the eigenenergies, $E^{5q}_\alpha$.  
   As discussed in the previous section \ref{sec:MBand5q}, 
   the matrix elements are assumed to be proportional to the spectroscopic factor, the overlap 
   $\Braket{ i | \alpha}$, 
   \be
   \Braket{ i | V | \alpha} = f \Braket{ i | \alpha}
   \label{eq:approximationV}
   \ee
where $f$ is the only parameter to determine the overall strength of the matrix elements.  
As we will discuss later, the approximation (\ref{eq:approximationV}) turns out to be rather good in comparison 
with the quark cluster model calculations~\cite{Takeuchi:2016ejt}.  

\begin{figure}[t]  
   \begin{center}
    \includegraphics[width=0.5\linewidth,clip]{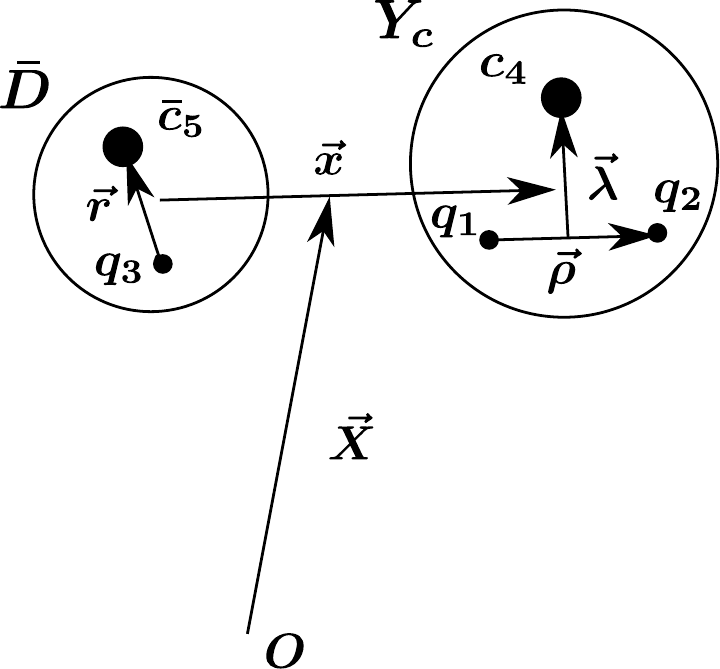}    
   \end{center}
  \caption{\label{fig:Jacobi_DYc}
  Jacobi coordinates of ``$\bar{D}$ meson'' and ``$Y_c$ baryon''
  in the $5q$ configuration.
  $q_i$ $(i=1,2,3)$ stands for the light quark, and $c_4$ ($\bar{c}_5$)
  stands for the (anti)charm quark.
  The coordinate $\vec{\rho}$ is the relative coordinate of $q_1q_2$,
  $\vec{\lambda}$ 
  the relative coordinate between the center of mass of
  $q_1q_2$ and $c_4$,
  $\vec{r}$ 
  the relative coordinate of $q_3\bar{c}_5$, and
  $\vec{x}$ 
  the relative coordinate between the centers of mass of
  $q_1q_2c_4$ and $\bar{c}_5q_3$.
  Though we do not use the total center-of-mass coordinate $\vec{X}$ in the present paper explicitly,
  it is also shown in the figure.
  }  
 \end{figure}

For the computation of the spectroscopic factor, let us construct the $MB$ and 5$q$ wave functions explicitly.  
We employ the standard non-relativistic  quark model with a harmonic oscillator confining potential. 
The wave functions are written as the products of color, spin, flavor and orbital wave functions.  
Let us introduce the notation 
$\Ket{\bar D Y_c (\vec{p}_i)}$ 
for
the open-charm meson-baryon channel $i$ 
of relative momentum 
$\vec{p}_i$.
Thus, we can write 
the wave function for 
$\Ket{\bar D Y_c (\vec{p}_i)}$
as~\cite{Hosaka:2004bn}
\be
 \Braket{ \vec \rho, \vec \lambda, \vec r, \vec x | \bar D Y_c (\vec{p}_i)}  =
    \psi^{int}_{\bar D}(\vec{r})   
    \psi^{int}_{Y_c}(\vec{\rho},\vec{\lambda})
   e^{i\vec p_i \cdot \vec x} 
    \times
    \phi_{\bar D Y_c }(CSF) .
    \label{eq:WF_MB}
\ee
In~\eqref{eq:WF_MB}, we indicate only the spatial 
coordinates explicitly, while the other coordinates 
for the color, spin and flavor 
are summarized in $\phi_{\bar{D} Y_c}(CSF)$.    
These
coordinates are shown in Fig.~\ref{fig:Jacobi_DYc}.  
The spatial wave functions 
$
\psi^{int}_{\bar D}(\vec{r}\,)   
    \psi^{int}_{\Lambda_c}(\vec{\rho},\vec{\lambda}\,)
$
are then written by 
those of
harmonic oscillator.

For 
the five-quark state,
we assume that the 
quarks 
move independently 
in a single confined region, and hence the $\vec x$ motion is also confined.  
Therefore, 
by introducing $\Ket{5q\, (\alpha)}$,
we have 
\be
\Braket{ \vec \rho, \vec \lambda, \vec r, \vec x | 5q\,(\alpha)}
=\psi^{int}_{5q}(\vec{\rho},\vec{\lambda},\vec{r})
    \left(\frac{2A}{\pi}\right)^{3/4}e^{-A^2 x^2}
       \times
    \phi_{5q}(CSF) ,
    \label{eq:WF_5q}
\ee
where 
the index $\alpha$ is for the $5q$ configurations, as shown in Table~\ref{table:5qchannels} for a given spin.
The parameter $A$ 
represents 
the inverse of the spatial separation of $\vec x$-motion, 
corresponding to the $qqc$ and $q \bar c$ clusters, which is 
in the 
order of 1 fm, or less.  
Again, the color, spin and flavor part is summarized in $\phi_{5q}(CSF)$. 

Now the spectroscopic factor is the overlap of \eqref{eq:WF_MB} and \eqref{eq:WF_5q}.  
Assuming that the spatial wave functions 
$
\psi^{int}_{\bar D}(\vec{r}\,)   
    \psi^{int}_{\Lambda_c}(\vec{\rho},\vec{\lambda}\,)
$
and 
$
\psi^{int}_{5q}(\vec{\rho},\vec{\lambda},\vec{r})
$ 
are the same, the overlap 
is given 
by
the color, spin and flavor parts, 
as labeled by $CSF$ below,
and 
by
the Fourie transform of the Gaussian function, 
\be
    \Braket{\bar{D}Y_c (\vec{p}_i)|5q\,(\alpha)}
    &=&
    \Braket{\phi_{\bar D Y_c} (CSF)| \phi_{5q} (CSF)}
    \int d^3 x
    \left(\frac{2A}{\pi}\right)^{3/4}e^{-A x^2}
    e^{i\vec{p_i}\cdot\vec{x}}
    \nonumber  \\
    &=&
    \Braket{ \phi_{\bar D Y_c} (CSF)| \phi_{5q} (CSF)}
    \left(\frac{2\pi}{A}\right)^{3/4} e^{-p^2_i/4A} 
    \equiv  S^\alpha_i g(\vec{p}_i),
\ee
where $S^\alpha_i$ is the spectroscopic factor for the color, flavor and
spin parts of the wave function, and $g(\vec p_i)$  the form factor for the transition 
$\bar D Y_c(\vec{p}_i) \to 5q(\alpha)$.  
The method how to compute $S^\alpha_i$ is presented in Appendix~\ref{sec:S-factor}, and the results for various
meson-baryon  channels $i$ and the $5q$ channels are summarized in Table~\ref{table:Sfactor}.  
  
\begin{table}[t]
 \caption{
 Spectroscopic factor of the $5q$ potential.
 $J$ is the total angular momentum of the system,
 $S_{c\bar{c}}$ is the total spin of $c\bar{c}$, and
 $S_{3q}$ is the total spin of the three light quarks.
 }
 \label{table:Sfactor}
 \begin{center}
  \begin{tabular}{ccc|cccccc}
   \hline\hline
   $J$&$S_{c\bar{c}}$&$S_{3q}$ &$\bar{D}\Lambda_{\rm c}$ &$\bar{D}^{\ast}\Lambda_{\rm c}$
   &$\bar{D}\Sigma_{\rm c}$ &$\bar{D}\Sigma^{\ast}_{\rm c}$
   &$\bar{D}^{\ast}\Sigma_{\rm c}$ &$\bar{D}^\ast\Sigma^\ast_{\rm c}$ \\ \hline
   $\frac{1}{2}$&$0$&$\frac{1}{2}$
	   &$0.35$ &$0.61$&$-0.35$ &--- &$0.20$  &$-0.58$ \\ 
   &$1$&$\frac{1}{2}$
	   &$0.61$ &$-0.35$ &$0.20$&--- &$-0.59$ &$-0.33$ \\
   &$1$&$\frac{3}{2}$
	   &$0.00$ &$0.00$ &$-0.82$ &---&$-0.47$  &$0.33$ \\ \hline
   $\frac{3}{2}$ &$0$&$\frac{3}{2}$
	   &--- &$0.00$ &--- &$-0.50$ &$0.58$  &$-0.65$ \\
   &$1$&$\frac{1}{2}$
	   &--- &$0.71$ &--- &$0.41$ &$-0.24$  &$-0.53$ \\
   &$1$&$\frac{3}{2}$
	   &--- &$0.00$&--- &$-0.65$&$-0.75$  &$-0.17$ \\ \hline
   $\frac{5}{2}$&$1$&$\frac{3}{2}$
	   &--- &--- &--- &--- &--- &$-1.00$ \\ 
   \hline\hline
  \end{tabular}
 \end{center}
\end{table}  

The wave functions should reflect the
  antisymmetric nature (a quark exchange effect)
  under the permutation among all light quarks especially in different clusters $\bar{D}Y_c$.
  This is neglected in $\Ket{\bar{D}Y_c (\vec{p}_i)}$.
  The effect, however, is introduced in the present model at least partially by considering the above overlap,
  because the $\psi^{int}_{5q}\phi_{5q}$ is totally antisymmetric over the quarks.
  Such quark exchange effect is suppressed,
  as the two color-singlet clusters $\bar{D}Y_c$ are
  further apart for larger $x$ and therefore the above overlap is suppressed.

  Finally, 
  the transition amplitude from $i$ to $j$ of 
  $\bar D Y_c$ channels
  is expressed by
  \begin{align}
   T_{ij} 
   = f^\prime\sum_\alpha S^\alpha_i S^\alpha_j g(\vec{p}_i) g(\vec{p}_j) \frac{1}{E-E^{5q}_\alpha}.
   \label{eq:Tij}
  \end{align}
  The overall strength $f^\prime$ of this amplitude is not determined, and is treated as a parameter,
  while the relative strengths of various channels $i,j$ are determined by the
  factors $S^\alpha_i$ and $S^\alpha_j$.
  
  The transition amplitude $T_{ij}$ in (\ref{eq:Tij}) has been given in a separable form. To use it in the
  Schr\"odinger equation, it is convenient to express it in the form of local potential, which
  is a function of the momentum transfer $\vec{q} = \vec{p}_i - \vec{p}_f$.
  We attempt to set
  \begin{align}
   g(\vec{p}_i)g(\vec{p}_j)=e^{-({p}^2_i+{p}^2_j)/4\alpha}\sim e^{-\beta{q}^2}.
  \end{align}
  On ignoring the angle-dependent term of 
  ${q}^2=(\vec{p}_i - \vec{p}_f)^2={p}^2_i+{p}^2_j-2\vec{p}_i\cdot\vec{p}_j$,
  it is reasonable to set
  $\beta=1/4A$.
  Therefore, the transition amplitude is parametrized as
  \begin{align}
   T_{ij} \sim \sum_\alpha S^\alpha_iS^\alpha_j
   e^{-q^2/4A}
   \frac{1}{E-E^\alpha_{5q}}.   
  \end{align}
  This gives an energy dependent local potential 
  \begin{align}
   V^{5q}_{ij}(E;r)\sim \sum_\alpha S^\alpha_iS^\alpha_j
   e^{-A r^2}
   \frac{1}{E-E^\alpha_{5q}},
  \end{align}
  with the relative coordinate $r$ between the heavy meson and baryon.
  
  Now, if we further expect that the compact five-quark configuration 
  $\Ket{5q\,(\alpha)}$
  is located sufficiently
  above the energy region in which we are interested, namely
  $E^\alpha_{5q} \gg m_{\bar{D}}+m_{Y_c} $,
  then we may further approximate
  \begin{align}
   V^{5q}_{ij}(r)=-f \sum_\alpha S^\alpha_iS^\alpha_j
   e^{-A r^2},
   \label{eq:Vshort_local}
  \end{align}
  where $f$ is a positive overall coupling strength.
  As shown in Table~\ref{table:E_5q}, in a simple quark model estimation,
  the $qqqc{\bar c}$ five-quark masses with the color-octet three light quarks are about 400 MeV
  larger than the threshold energies of $\bar{D} Y_c$ in the present study.
  The masses of hidden-bottom five-quarks are similarly higher than the $\bar{B} Y_b$ thresholds.
  This makes the potential~\eqref{eq:Vshort_local} attractive for both of the hidden-charm and hidden-bottom sectors.
  As we will discuss later in this paper,
  especially this attraction turns out to be
  the driving force for abundant $P_c$ states.
  
\begin{table}[htbp]
\caption{
 Masses of the hidden-charm five-quark states with the color-octet three light quarks, $E^\alpha_{5q}$,
calculated by using parameters in Ref.~\cite{Takeuchi:2016ejt}.
All the entries are listed in MeV.
$J$ stands for the total spin of the five-quarks, [$q^38s]S$ stands for 
the five-quark state, which consists of the $uud$ quarks with 
a spin of $s$ and the $c\overline{c}$ pair with a spin of $S$.   
\label{table:E_5q}
}
\renewcommand\arraystretch{1.5}
\begin{center}
\begin{tabular}{ccccccccccccc} \hline
$J$&
$[q^3 8\frac{1}{2}]0$ &
$[q^3 8\frac{1}{2}]1$ &
$[q^3 8\frac{3}{2}]0$ & 
 $[q^3 8\frac{3}{2}]1$ \\ \hline
$\frac{1}{2}$ &
4816.2 &
4759.1 & - &
4772.2 \\
$\frac{3}{2}$ 
 & 
- &
4822.3 &
4892.5 &
4835.4 & 
\\
 $\frac{5}{2}$
 & - & - & - &
4940.7
\\ \hline
\end{tabular}
\end{center}
\label{tbl2}
\end{table}%

  \section{Numerical results}
  \label{sec:results}

  \subsection{Model parameters}
  \label{sec:Model_parameters}

To start with, let us fix the two parameters, $f$ and $A$, in the $5q$ potential~\eqref{eq:Vshort_local}.  
The Gaussian range $A=\mu\omega/2$ originates the frequency of the harmonic oscillator
potential $V(x)=\frac{1}{2}\mu\omega^2x^2$
of a ``meson'' and a ``baryon'' in the $5q$ state, as shown in Fig.~\ref{fig:Jacobi_DYc}.
Hence, $A$ is expressed by the relative distance
$\Braket{x^2}\equiv\Braket{\psi|x^2|\psi}$ of the ``meson'' and
``baryon'' as 
\begin{align}
 A=\frac{3}{4\Braket{x^2}}, 
\end{align}
with
the harmonic oscillator wave function 
\begin{align}
 &\psi(x)=\left(\frac{2A}{\pi}\right)^{3/4} e^{-A x^2} \, .
\end{align}
In this study, we assume that $\sqrt{\Braket{x^2}}$ is less than 1~fm,
namely $A\geq \frac{3}{4}$ fm$^{-2}$, and employ $A=1$ fm$^{-2}$.

The overall strength $f$ is a free parameter, 
and we will show our numerical results for various $f$.
It is then convenient to set a reference value $f_0$.  
Here we use the $\bar{D}^\ast \Sigma_{\rm c}$ diagonal term of the OPEP, 
\begin{align}
 f_0=
 \left|C^{\pi}_{\bar{D}^{\ast}\Sigma_{\rm c}}(r=0)\right|\sim 6 \text{ MeV},
\end{align}
where $C^{\pi}_{\bar{D}^{\ast}\Sigma_{\rm c}}(r)\equiv -\frac{gg_1}{3f_\pi^2}C(r)$ 
is the central force of 
$V^{\pi}_{\bar{D}^{\ast}\Sigma_{\rm c}-\bar{D}^{\ast}\Sigma_{\rm c}}(r)$ 
without the spin-dependent operator $\vec{S}\cdot\vec{\sigma}$, as shown
in Eq.~\eqref{eq:OPEP-P*B6-P*B6}.

\begin{figure}[t]
  \begin{center}
   \includegraphics[width=0.5\linewidth,clip]{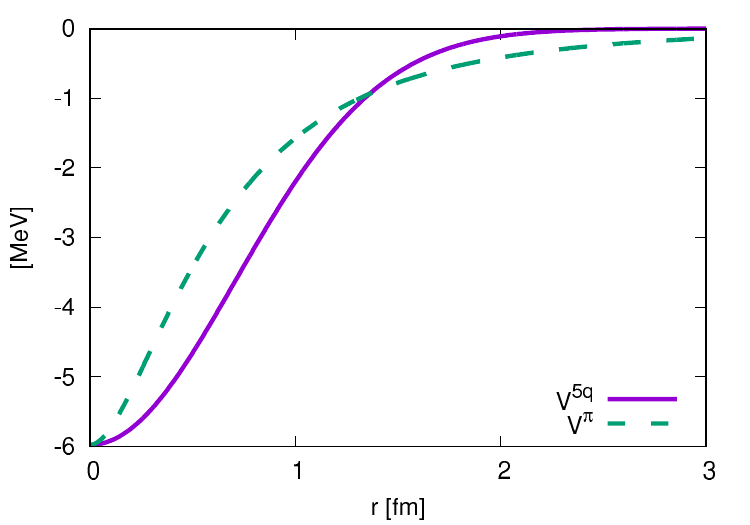}   
   \caption{\label{fig:bare_potential}
   (Color online)
   The plot of the $5q$ potential, $V^{5q}$, (solid line) and the central force of the
   OPEP in the diagonal
   $\bar{D}^\ast\Sigma_{\rm c}-\bar{D}^\ast\Sigma_{\rm c}$ term,
   $V^{\pi}$, (dashed line).
   }  
  \end{center}
\end{figure}

When $f_0 = 6$ MeV and $A = 1$ fm$^{-2}$ are used, 
the short range interaction is not as strong as what we expect 
from the $NN$ force.  
To see this point, we compare the volume  
integrals of the potentials~\footnote{
The volume integral corresponds to the potential in the momentum space
at zero momentum.
Therefore, it makes an important contribution to the amplitude in the low-energy scattering. 
}
 \begin{align}
  &
  \left|\int d^3rf_0e^{-A r^2}\right|
  =
  4.3\times 10^{-6}
  \text{ MeV}^{-2}, \label{eq:volume_int_5q} \\
 &
  \left|\int d^3rC^\pi_{\bar{D}^\ast\Sigma_{\rm c}}(r)\right|
  =
  1.8\times 10^{-5}
  \text{ MeV}^{-2}, \label{eq:volume_int_OPEP} \\
 &
  \left|\int d^3rV^\pi_{NN}(r)\right|
  =
  6.3\times 10^{-5}  
  \text{ MeV}^{-2}, \label{eq:volume_int_NNBonn} \\
  &\left|\int d^3rV^\sigma_{NN}(r)\right|
  =
  3.8\times 10^{-3}
  \text{ MeV}^{-2}, \label{eq:volume_int_NNsigBonn} 
 \end{align}
 with
 the central force of the OPEP and the 
 $\sigma$ 
 exchange, $V^{\pi}_{NN}$
 and $V^{\sigma}_{NN}$, in the Bonn potential~\cite{Machleidt:1987hj}.
 From Eqs.~\eqref{eq:volume_int_5q}-\eqref{eq:volume_int_NNsigBonn},
 we obtain
 \begin{align}
  \left|\int d^3rf_0 e^{-A r^2}(r)\right|\sim\frac{1}{4}\left|\int
  d^3r C^\pi_{\bar{D}^\ast\Sigma_{\rm c}}(r)\right|
  \sim\frac{1}{15}\left|\int d^3rV^\pi_{NN}(r)\right|\sim\frac{1}{880}\left|\int d^3rV^\sigma_{NN}(r)\right|.
 \end{align}
 We find that the volume integral of the $5q$
 potential with $f=f_0$~\eqref{eq:volume_int_5q}
 is smaller than that of the $NN$
 potentials~\eqref{eq:volume_int_NNBonn} and
 \eqref{eq:volume_int_NNsigBonn}.
 In particular, the volume integral in Eq.~\eqref{eq:volume_int_5q} is much smaller than
 in Eq.~\eqref{eq:volume_int_NNsigBonn} 
 for 
 the $\sigma$ exchange
 potential
 in the $NN$ interaction.
 In Section~\ref{sec:results}, 
 we will see that
 the non-trivial bound and resonant states are produced,
when 
$f\sim 25f_0$ (or larger), whose volume integral is still
 much smaller than that in Eq.~\eqref{eq:volume_int_NNsigBonn}.
In Fig.~\ref{fig:bare_potential},
we show the $5q$ potential with the fixed parameters $f_0$ and $A$, where
the obtained $5q$ potential is compared with
$C^\pi_{\bar{D}^\ast\Sigma_{\rm c}}(r)$.

  \subsection{Numerical methods}
  \label{sec:Numerical_methods}
The bound and resonant states are obtained by solving the
coupled-channel Schr\"odinger equation with the OPEP, $V^{\pi}(r)$, and
$5q$
potential, $V^{5q}(r)$, 
\begin{align}
 \left(K+V^{\pi}(r)+
 V^{5q}(r)
 \right)\Psi(r)=E\Psi(r),
 \label{eq:Shrodinger_eq}
\end{align}
with the kinetic term $K$.
The OPEP and kinetic terms are summarized in Appendix~\ref{sec:appendix_OPEP}.

The Schr\"odinger equation~\eqref{eq:Shrodinger_eq} is solved by using the variational method.
The trial function $\Psi_{JM,IM_I}(\vec{r}\,)$ with the total angular
momentum $J$, total isospin $I$, and their $z$-components 
$M$ and $M_I$ is expressed by the Gaussian expansion method~\cite{Hiyama:2003cu} as
\begin{align}
 \Psi_{JM,IM_I}(\vec{r}\,)&=\sum_{i=1}^{i_{max}}\sum_{L,S}C_{iLS} \left[\psi_{iLM_L}(\vec{r}\,)\otimes \left[\chi_{s_1m_{s_1}}\chi_{s_2m_{s_2}}\right]_{SM_S}\right]_{JM}
 \left[\eta_{I_1m_{I_1}}\eta_{I_2m_{I_2}}\right]_{IM_I}, \\
 \psi_{iLM_L}(\vec{r}\,)&=\sqrt{\frac{2}{\Gamma (L+3/2)b^3_i}}\left(\frac{r}{b_i}\right)^{L}
 \exp\left(-\frac{r^2}{2b^2_i}\right)Y_{LM_L}(\hat{r}).
 \label{eq:GEM}
\end{align}
In the Gaussian expansion method, the wave function is expanded in terms of Gaussian basis functions, 
as shown in Eq.~\eqref{eq:GEM}.
The coefficients $C_{iLS}$ are determined by diagonalizing the Hamiltonian, and
$\psi_{iLM_L}(\vec{r}\,)$ are the radial wave function of the
meson-baryon with the orbital angular momentum $L$
and the $z$-component $M_L$. 
The (iso)spin wave functions $\chi_{s_km_{s_k}}$ ($\eta_{I_km_{I_k}}$) with $k=1,2$
are 
for the (iso)spin $s_k$ ($I_k$) of the hadron $k$,
with
the $z$-component 
$m_{s_k}$ ($m_{I_k}$).
The total (iso)spin is given by
$S$ ($I$) 
with the $z$-component $M_S$ ($M_I$). 
The angular part of the radial wave function is represented by 
the spherical harmonics $Y_{LM_L}(\hat{r})$.
The Gaussian ranges $b_i$ are given by the form of geometric series as
\begin{align}
 b_i=b_1 a^{i-1} \quad (i=1,\cdots, i_{max}),
\end{align}
with the variational parameters $b_1$ and $b_{i_{max}}$, and $a= (b_{i_{max}}/b_1)^{1/({i_{max}-1})}$.

In order to find not only bound states, but also resonances,
the complex scaling
method~\cite{Aguilar1971_269,Blaslev1971_22,Simon1972_27,Prog.Theor.Phys11620061Aoyama}
is employed.
By diagonalizing the complex scaled Hamiltonian with $r\rightarrow re^{i\theta}$ and
$p\rightarrow pe^{-i\theta}$,
binding energies and resonance energies with decay widths are
obtained as the eigenenergy of the complex scaled Schr\"odinger equation.

  \subsection{Numerical results of the hidden-charm sector}
  \label{sec:Numerical_results_charm}
\begin{figure}[t]
 \begin{center}
    \begin{tabular}{ccc}
     $J^P=1/2^-$&&\\
     (i) $(S_{c\bar{c}},S_{3q})=(0,\frac{1}{2})$ 
     &	(ii) $(S_{c\bar{c}},S_{3q})=(1,\frac{1}{2})$ 
	 &  (iii) $(S_{c\bar{c}},S_{3q})=(1,\frac{3}{2})$ \\ 
   \includegraphics[width=0.34\linewidth,clip]{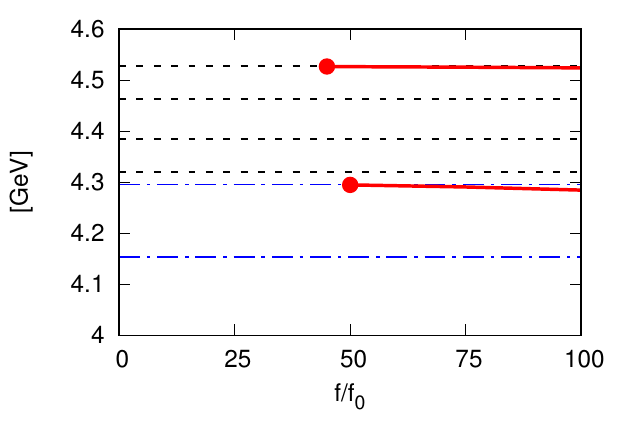}&
   \includegraphics[width=0.34\linewidth,clip]{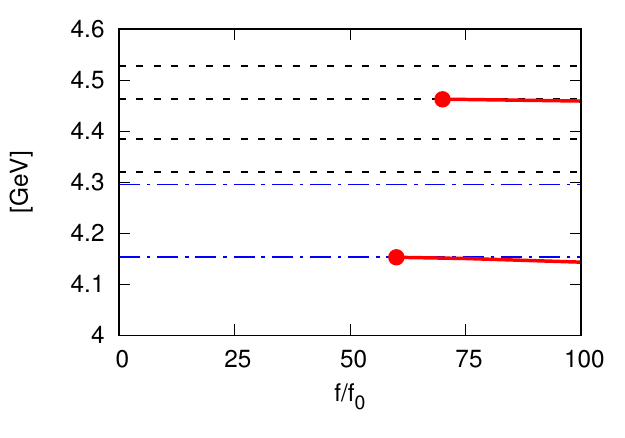}&
   \includegraphics[width=0.34\linewidth,clip]{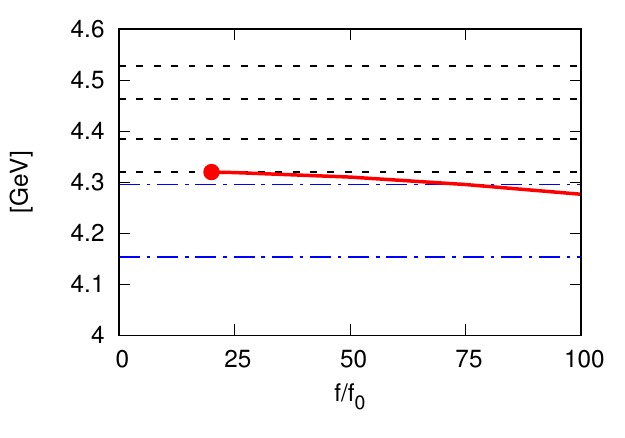}\\
    \end{tabular}  
  \caption{\label{fig:results_12m}
  (Color online) Bound and resonant state energies of the hidden-charm
  molecules (solid lines)
  with various coupling constants
  $f$ for $J^P=1/2^-$, using 
  the OPEP and one of the three $5q$ potentials derived from the configuration 
  (i) $(S_{c\bar{c}},S_{3q})=(0,1/2)$,
  (ii) $(1,1/2)$, 
  or 
  (iii) $(1,3/2)$.
  The horizontal axis shows the ratio $f/f_0$, where $f_0$ is the
  reference value defined in Sec.~\ref{sec:Model_parameters}.
  Filled circle is the starting point where the states appear.
  Dashed lines are the $\bar{D}\Sigma_{\rm c}$, $\bar{D}\Sigma^{\ast}_{\rm c}$,
  $\bar{D}^{\ast}\Sigma_{\rm c}$, and $\bar{D}^{\ast}\Sigma^{\ast}_{\rm c}$
  thresholds.
  Dot-dashed lines are the $\bar{D}\Lambda_{\rm c}$ and
  $\bar{D}^{\ast}\Lambda_{\rm c}$ thresholds.
  }  
 \end{center}
\end{figure}
\begin{figure}[t]
   \begin{center}
    $J^P=1/2^-$\\
   \includegraphics[width=0.6\linewidth,clip]{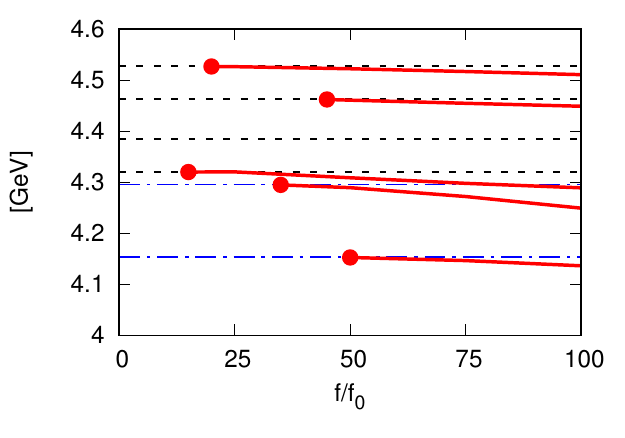}    
    \caption{\label{fig:results_12m_Sum}
    (Color online)
    The same as Fig.~\ref{fig:results_12m} for the bound and resonant
    states of the hidden-charm molecules for $J^P=1/2^-$ using the
    OPEP and the sum of the three $5q$ potentials.
    }
   \end{center}
\end{figure}

\begin{figure}[t]
 \begin{center}
    \begin{tabular}{ccc}
     $J^P=3/2^-$&&\\
    (i) $(S_{c\bar{c}},S_{3q})=(0,\frac{3}{2})$ 
    &(ii) $(S_{c\bar{c}},S_{3q})=(1,\frac{1}{2})$ 
	& (iii) $(S_{c\bar{c}},S_{3q})=(1,\frac{3}{2})$ \\ 
   \includegraphics[width=0.34\linewidth,clip]{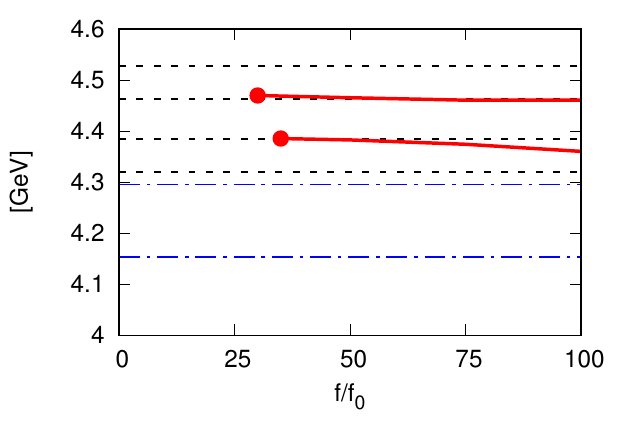}&
   \includegraphics[width=0.34\linewidth,clip]{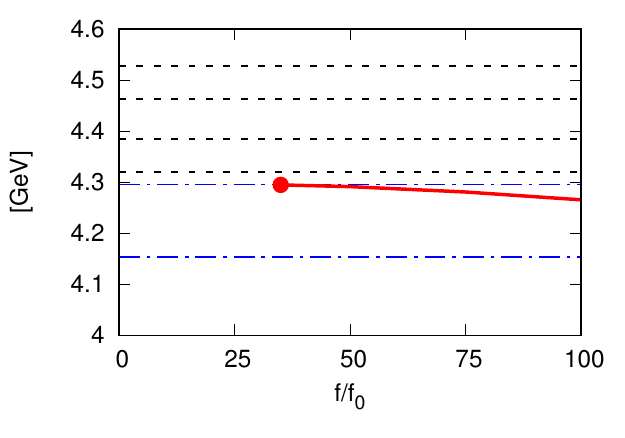}&
   \includegraphics[width=0.34\linewidth,clip]{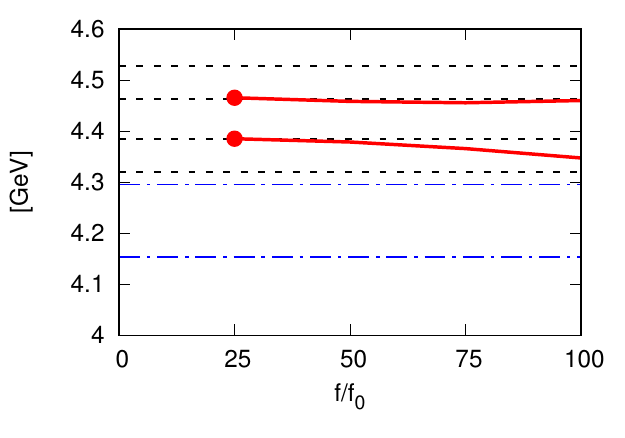}\\
    \end{tabular}  
  \caption{\label{fig:results_32m} 
  (Color online)
  The same as Fig.~\ref{fig:results_12m} for the resonant
  states of the hidden-charm molecules for $J^P=3/2^-$ using the
  OPEP and 
  one of the three $5q$ potentials derived from the configuration 
  (i)~$(S_{c\bar{c}},S_{3q})=(0,3/2)$,
  (ii)~$(1,1/2)$, or (iii)~$(1,3/2)$.
  }  
 \end{center}
\end{figure}

\begin{figure}[t]
   \begin{center}
    $J^P=3/2^-$\\
   \includegraphics[width=0.6\linewidth,clip]{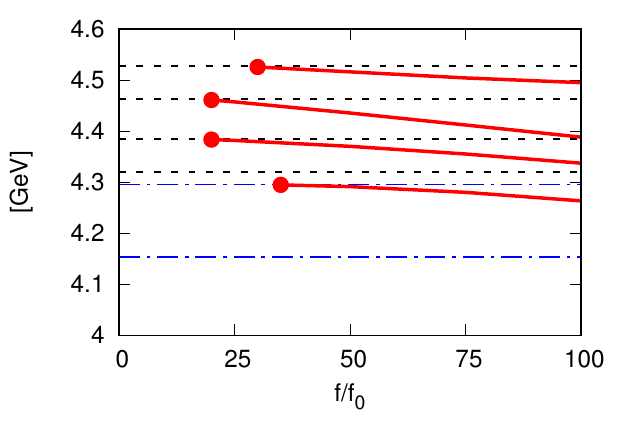}
    \caption{\label{fig:results_32m_Sum}
    (Color online)
    The same as Fig.~\ref{fig:results_12m} for the resonant
    states of the hidden-charm molecules for $J^P=3/2^-$ using the
    OPEP and the sum of the three $5q$ potentials.
    }  
   \end{center}
\end{figure}

\begin{figure}[t]
  \begin{center}
   \begin{tabular}{c}    
     $J^P=5/2^-$ \\ 
    \includegraphics[width=0.6\linewidth,clip]{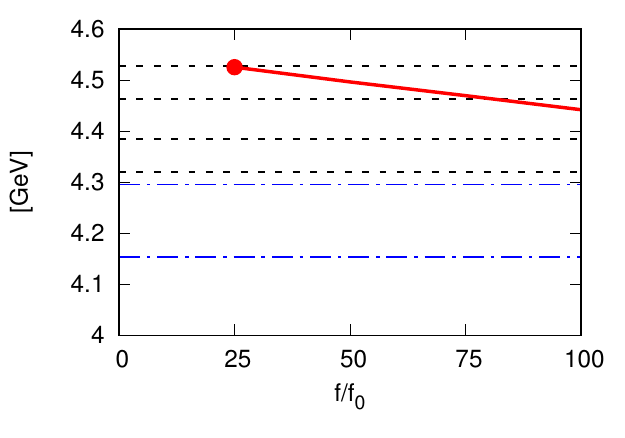}   \\
   \end{tabular}
   \caption{\label{fig:results_52m}
   (Color online)
   The same as Fig.~\ref{fig:results_12m} for the resonant
   states of the hidden-charm molecules for $J^P=5/2^-$ using the
   OPEP and the $5q$ potential from the 
   configuration $(S_{c\bar{c}},S_{3q})=(1,\frac{3}{2})$.
  }  
  \end{center}
\end{figure}

Let us show the numerical results of the hidden-charm meson-baryon
molecules.
The coupling strength $f$ dependence of the energy spectrum is
summarized 
in Figs.~\ref{fig:results_12m}-\ref{fig:results_12m_Sum}
and Tables~\ref{table:energyspectra_12m}-\ref{table:energyspectra_12m_SUM}
for $J^P=1/2^-$,
in Figs.~\ref{fig:results_32m}-\ref{fig:results_32m_Sum} and
Tables~\ref{table:energyspectra_32m}-\ref{table:energyspectra_32m_SUM} for
$J^P=3/2^-$, and
in Fig.~\ref{fig:results_52m} and Table~\ref{table:energyspectra_52m} for $J^P=5/2^-$.

Figure~\ref{fig:results_12m} shows the strength $f$ dependence of the
obtained energy spectra for $J^P=1/2^-$ by employing the OPEP and
one of the three $5q$ potentials derived from the configurations 
(i)
$(S_{c\bar{c}},S_{3q})=(0,1/2)$, (ii) $(1,1/2)$, or (iii) $(1,3/2)$.
We obtain no state only with the OPEP, corresponding to the result at $f/f_0=0$, while
the bound and resonant states appear by increasing the strength $f$ of
the $5q$ potential.
The filled circle in figures 
shows the starting point where the state is found.
In Fig.~\ref{fig:results_12m} (i), two resonances appear below
$\bar{D}^\ast\Lambda_{\rm c}$ and $\bar{D}^\ast \Sigma^\ast_{\rm c}$
thresholds 
for $f$ larger than $f/f_0=50$ and $45$, respectively.
In Fig.~\ref{fig:results_12m} (ii), the bound state and resonance are obtained below
$\bar{D}\Lambda_{\rm c}$ and $\bar{D}^\ast \Sigma_{\rm c}$
thresholds 
for $f$ larger than
$f/f_0=60$ and $70$, respectively.
In Fig.~\ref{fig:results_12m} (iii),
the resonance below the $\bar{D}\Sigma_{\rm c}$ threshold appears 
at and above
$f/f_0=20$ which is smaller 
than the strength in other channels.
Thus, the $5q$ potential from the configuration with $S_{3q}={3}/{2}$ produces the strong attraction rather 
than the potential from the configuration with $S_{3q}={1}/{2}$, corresponding to the results in Figs.~\ref{fig:results_12m} (i) and (ii).

As shown in Fig.~\ref{fig:results_12m},
the energy spectra 
appear just below the meson-baryon thresholds.
The obtained spectrum structure can be explained by the
spectroscopic factor ($S$-factor) of the $5q$ potential in
Table~\ref{table:Sfactor}.
Since the 
$S$-factor
gives the relative strength of the
$5q$
potential
among $\bar{D}^{(\ast)}\Lambda_{\rm c}$ and
$\bar{D}^{(\ast)}\Sigma^{(\ast)}_{\rm c}$ channels,
the channels with a large
$S$-factor
play an
important role to produce 
bound and resonant states.
For (i) $(S_{c\bar{c}},S_{3q})=(0,1/2)$,
the large $S$-factors are obtained for 
the $\bar{D}^\ast \Lambda_{\rm c}$ and $\bar{D}^\ast \Sigma^\ast_{\rm c}$
channels 
and
indeed, the resonances are obtained below the $\bar{D}^\ast \Lambda_{\rm c}$ and
$\bar{D}^\ast \Sigma^\ast_{\rm c}$ thresholds.
In (ii) $(S_{c\bar{c}},S_{3q})=(1,1/2)$,
the bound and resonant states below $\bar{D}\Lambda_{\rm c}$ and
$\bar{D}^\ast\Sigma_{\rm c}$ are obtained, where the large $S$-factors
are obtained in the $\bar{D}\Lambda_{\rm c}$ and $\bar{D}^\ast\Sigma_{\rm c}$ channels.
In (iii) $(S_{c\bar{c}},S_{3q})=(1,3/2)$, one resonance below the
$\bar{D}\Sigma_{\rm c}$ threshold is found, where
the large $S$-factor is obtained in the $\bar{D}\Sigma_{\rm c}$ channel.

In Fig.~\ref{fig:results_12m_Sum}, we show the energy spectra with the
full potential including
OPEP and the sum of the three $5q$ potentials with the same weight.
As expected, the result is a combination of the three results in
Fig.~\ref{fig:results_12m} with some more attraction.
As $f/f_0$ is increased, the resonance appear even for $f/f_0=15$, 
which would corresponds to
the state found in Fig.~\ref{fig:results_12m} (iii).
We see that the $5q$ potential
produces many states 
when the strength $f/f_0$ is increased.


The states are also obtained in $J^P=3/2^-$ and $5/2^-$ as well as
$1/2^-$, where the structure of the energy spectra is explained by the $S$-factor.
In Figs.~\ref{fig:results_32m} and \ref{fig:results_32m_Sum},
the strength $f$ dependence of the energies for $J^P=3/2^-$ is shown.
We also obtain no state only with the OPEP, corresponding to the results at $f/f_0=0$, 
but the states appear when the strength of the $5q$ potential is increased as seen in $J^P=1/2^-$.
There are three $5q$ potentials derived from the quark configurations (i) $(S_{c\bar{c}},S_{3q})=(0,3/2)$,
(ii) $(1,1/2)$, and (iii) $(1,3/2)$.
In Fig.~\ref{fig:results_32m} (i), 
two resonances are obtained 
near the $\bar{D}\Sigma^{\ast}_{\rm c}$ and
$\bar{D}^\ast\Sigma_{\rm c}$ thresholds, 
where 
the large $S$-factors are obtained in the $\bar{D}\Sigma^\ast_{\rm c}$,
$\bar{D}^\ast\Sigma_{\rm c}$, and $\bar{D}^\ast\Sigma^\ast_{\rm c}$ components.
In Fig.~\ref{fig:results_32m} (ii),
one resonance is found near the $\bar{D}^{\ast}\Lambda_{\rm c}$
threshold for $f/f_0\geq35$, where the $S$-factor of the $\bar{D}^{\ast}\Lambda_{\rm c}$ is also large.
In Fig.~\ref{fig:results_32m} (iii),
the two resonances are found near the $\bar{D}\Sigma^{\ast}_{\rm c}$ and
$\bar{D}^\ast\Sigma_{\rm c}$ thresholds, and the large $S$-factors are also obtained 
in the $\bar{D}\Sigma^{\ast}_{\rm c}$ and $\bar{D}^\ast\Sigma_{\rm c}$ channels.
In Fig.~\ref{fig:results_32m_Sum},
the results with the summation of the three $5q$ potentials are shown.
The four resonances appear
below the $\bar{D}\Lambda^\ast_{\rm c}$ threshold for $f/f_0\geq35$,
below the $\bar{D}\Sigma^\ast_{\rm c}$ threshold for $f/f_0\geq20$,
below the $\bar{D}^\ast\Sigma_{\rm c}$ threshold for $f/f_0\geq20$,
and below the $\bar{D}^\ast\Sigma^\ast_{\rm c}$ threshold for
$f/f_0\geq30$, respectively.

The obtained energy spectra 
for $J^P=5/2-$
are shown in Fig.~\ref{fig:results_52m}.
There is only one $5q$ potential
from the quark configuration 
$(S_{c\bar{c}},S_{3q})=(1,3/2)$, which appears only in the $\bar{D}^\ast\Sigma^\ast_{\rm c}$ channel.
No state is found only by employing the OPEP, while
one resonance below the $\bar{D}^\ast\Sigma^\ast_{\rm c}$ threshold
is obtained for $f/f_0\geq 25$.

The obtained results in the hidden-charm sector should be compared to
the $P^+_{\rm c}$ pentaquarks.
The LHCb collaboration reported that the two $P^+_{\rm c}$ pentaquarks were
found close to the $\bar{D}\Sigma^\ast_{\rm c}$ and $\bar{D}^\ast\Sigma_{\rm c}$ thresholds,
and the preferred spins are $J=3/2$ and $5/2$.
In the numerical results,
we also obtain the resonances close to the $\bar{D}\Sigma^\ast_{\rm c}$ and
$\bar{D}^\ast\Sigma_{\rm c}$ thresholds for $J^P=3/2^-$, as shown in
Figs.~\ref{fig:results_32m}-\ref{fig:results_32m_Sum},
and Tables~\ref{table:energyspectra_32m}-\ref{table:energyspectra_32m_SUM}.
The obtained resonances close to the $\bar{D}^\ast\Sigma_{\rm c}$
have the mass around 4460 MeV and the width around 20 MeV,
and these values are in good agreement with the observed $P^+_{\rm c}$,
while the spin-parity of the obtained state $J^P=3/2^-$ is not the
suggested one by the LHCb collaboration.
For the resonance close 
to
the $\bar{D}\Sigma^\ast_{\rm c}$ threshold,
the obtained mass around 4380 MeV 
agrees with 
the reported $P^+_{\rm c}$
mass.
However, the obtained width around 6 MeV is 
very different 
from the
reported width 205 MeV.
In comparison to the observed $P^+_{\rm c}$ states,
the $J^P=3/2^-$ state could be a candidate of the upper $P^+_{\rm c}$
state.


\begin{table}[htbp]
  \caption{\label{table:energyspectra_12m}
 Energy spectra of the hidden-charm 
 molecules for $J^P=1/2^-$
 using the OPEP and one of the $5q$ potentials from 
 the configuration (i) $(S_{c\bar{c}},S_{3q})=(0,1/2)$,
 (ii) $(1,1/2)$, or (iii) $(1,3/2)$.
 The energy $E$ and half decay width $\Gamma/2$ in the various coupling constants $f/f_0$ are shown.
 The third row is for the point, where the state appears.
 The fourth, fifth, sixth and seventh rows show the obtained values 
 with $f=25f_0$, $50f_0$, $75f_0$, and $100f_0$, respectively.
 The values are given in units of MeV.
 The lowest threshold $\bar{D}\Lambda_{\rm c}$ is at 4153.46 MeV,
 and the state whose energy is lower than the threshold is a bound state.} 
 \begin{center}
   \begin{tabular}{cc||c|cccc}
    \hline\hline    
    (i) $(0,1/2)$ &$f/f_0$&45 &25 &50 &75 &100 \\
    &$E$ [MeV]
	&4527 &--- &4527 &4526 &4524 \\ 
    &$\Gamma/2$ [MeV]&0.87 &--- &0.98 &1.77 &2.53 \\ \hline
    &$f/f_0$&50 &25 &50 &75 &100 \\
    &$E$ [MeV]
	&4295 &--- &4295 &4291 &4285 \\ 
    &$\Gamma/2$ [MeV]&0.22 &--- &0.22 &1.42 &4.33 \\ \hline\hline    
    (ii) $(1,1/2)$ & $f/f_0$&70 &25 &50 &75 &100 \\
    &$E$ [MeV]&4463 &--- &--- &4462 &4459 \\
    &$\Gamma/2$ [MeV]&1.44 &--- &--- &1.66 &2.37 \\ \hline
    &$f/f_0$&60 &--- &--- &75 &100 \\
    &$E$ [MeV]&4153 &--- &--- &4151 &4144 \\
    &$\Gamma/2$ [MeV]&--- &--- &--- &--- &--- \\ \hline\hline    
    (iii) $(1,3/2)$ & $f/f_0$&20 &25 &50 &75 &100 \\
    &$E$ [MeV]&4320 &4319 &4310 &4295 &4276 \\
    &$\Gamma/2$ [MeV]&0.33 &0.35 &0.15 &$3.90\times 10^{-3}$ &$8.21\times 10^{-2}$\\ 
    \hline\hline
   \end{tabular}
 \end{center}
\end{table}

\begin{table}[htbp]
  \caption{ \label{table:energyspectra_12m_SUM}
 The same as Table~\ref{table:energyspectra_12m} for the
 energy spectra of the hidden-charm molecules for
 $J^P=1/2^-$ using the OPEP and  the sum of the three $5q$ potentials.
 }
 \begin{center}  
  \begin{tabular}{cc||c|cccc}
   \hline\hline
   SUM&$f/f_0$ &20 &25 &50 &75 &100 \\
   &$E$ [MeV]
       &4527 &4526 &4523 &4517 &4511 \\ 
   &$\Gamma/2$ [MeV]
       &0.63 &0.85 &2.00 &2.79 &3.33 \\ \hline
   &$f/f_0$ &45 &25 &50 &75 &100 \\
   &$E$ [MeV]
       &4462 &--- &4461 &4455 &4449 \\ 
   &$\Gamma/2$ [MeV]
       &3.27 &--- &3.93 &6.54 &8.66 \\ \hline
   &$f/f_0$ &15 &25 &50 &75 &100 \\
   &$E$ [MeV]
       &4320 &4320 &4309 &4298 &4289 \\ 
   &$\Gamma/2$ [MeV]
       &0.45 &1.70 &3.40 &2.34 &2.57$\times 10^{-2}$ \\ \hline
   &$f/f_0$ &35 &25 &50 &75 &100 \\
   &$E$ [MeV]
       &4295 &--- &4290 &4272 &4249 \\
   &$\Gamma/2$ [MeV]
       &2.01$\times 10^{-2}$ &---  &6.17$\times 10^{-2}$ &9.23$\times 10^{-2}$ &7.93$\times 10^{-2}$ \\ \hline
      &$f/f_0$ &50 &25 &50 &75 &100 \\
   &$E$ [MeV]
       &4153 &--- &4153 &4147 &4136 \\ 
   &$\Gamma/2$ [MeV]
       &--- &--- &--- &--- &--- \\ \hline\hline
  \end{tabular}
 \end{center}
 \end{table}

\begin{table}[t]
  \caption{\label{table:energyspectra_32m}
 The same as Table~\ref{table:energyspectra_12m} for the
 energy spectra of the hidden-charm molecules for 
 $J^P=3/2^-$ using the OPEP and one of the three $5q$ potentials from the
 configuration
 (i)~$(S_{c\bar{c}},S_{3q})=(0,3/2)$,
 (ii)~$(1,1/2)$, or (iii)~$(1,3/2)$.}
  \begin{center}
    \begin{tabular}{cc||c|cccc}
     \hline\hline
    (i) $(0,3/2)$& $f/f_0$
     &30 &25 &50 &75 &100 \\
     &$E$ [MeV]
	 &4470 &--- &4466 &4461 &4461 \\ 
     &$\Gamma/2$ [MeV]
	 &10.49 &---& 17.16 &26.61 &38.75 \\ \hline
     & $f/f_0$
     &35 &25 &50 &75 &100 \\
     &$E$ [MeV]
	 &4386 &--- &4383 &4374 &4360 \\ 
     &$\Gamma/2$ [MeV]
	 &2.21 &--- &3.33 &4.08 &3.66 \\ \hline\hline
    (ii) $(1,1/2)$& $f/f_0$
     &35 &25 &50 &75 &100 \\
     &$E$ [MeV]
	 &4295 &--- &4292 &4281 &4265 \\ 
     &$\Gamma/2$ [MeV]
	 &2.64$\times 10^{-2}$ &--- &4.47$\times 10^{-2}$ &8.92$\times 10^{-4}$ &0.109 \\ \hline\hline
    (iii) $(1,3/2)$& $f/f_0$
     &25 &25 &50 &75 &100 \\
     &$E$ [MeV]
	 &4466 &4466 &4459 &4456 &4460 \\
     &$\Gamma/2$ [MeV]
	 &9.96 &9.96 &16.51 &23.50 &28.94 \\ \hline
     & $f/f_0$
     &25 &25 &50 &75 &100 \\
     &$E$ [MeV]
	 &4385 &4385 &4379 &4366 &4348 \\
     &$\Gamma/2$ [MeV]
	 &1.85 &1.85 &2.96 &2.45 &1.57 \\     
     \hline\hline
    \end{tabular}
  \end{center}
 \end{table}

\begin{table}[t]
  \caption{The same as Table~\ref{table:energyspectra_12m} for the
 energy spectra of the hidden-charm molecules for
 $J^P=3/2^-$ using the OPEP and the sum of the three $5q$ potentials.}
 \label{table:energyspectra_32m_SUM}
  \begin{center}
    \begin{tabular}{cc||c|cccc}
     \hline\hline
    SUM & $f/f_0$
     &30 &25 &50 &75 &100 \\
     &$E$ [MeV]
	 &4526 &--- &4516 &4505 &4495 \\ 
     &$\Gamma/2$ [MeV]
	 &9.58 &--- &13.52 &17.60 &22.34 \\ \hline
     & $f/f_0$
     &20 &25 &50 &75 &100 \\
     &$E$ [MeV]
	 &4461 &4457 &4436 &4412 &4389 \\ 
     &$\Gamma/2$ [MeV]
	 &11.61 &12.83 &14.70 &13.17 &10.56 \\ \hline
     & $f/f_0$
     &20 &25 &50 &75 &100 \\     
     &$E$ [MeV]
	 &4384 &4382 &4370 &4355 &4338 \\ 
     &$\Gamma/2$ [MeV]
	 &3.11 &3.62 &4.69 &4.86 &4.59 \\ \hline
     & $f/f_0$
     &35 &25 &50 &75 &100 \\
     &$E$ [MeV]
	 &4295 &---  &4291 &4280 &4264 \\
     &$\Gamma/2$ [MeV]
	 &1.41$\times 10^{-2}$ &--- &5.09$\times 10^{-2}$
		 &7.71$\times 10^{-2}$ &8.15$\times 10^{-2}$ \\     
     \hline\hline
    \end{tabular}
  \end{center}
 \end{table}

\begin{table}[t]
  \caption{\label{table:energyspectra_52m}
 The same as Table~\ref{table:energyspectra_12m} for the
 energy spectra of the hidden-charm molecules for
 $J^P=5/2^-$ using the OPEP and the $5q$ potential from the configuration $(S_{c\bar{c}},S_{3q})=(1,3/2)$.
 } 
  \begin{center}
    \begin{tabular}{cc||c|ccccc}
     \hline\hline
    $(1,3/2)$ & $f/f_0$
     &25 &25 &50 &75 &100 \\
     &$E$ [MeV]
	 &4526 &4526 &4496 &4470 &4442 \\ 
     &$\Gamma/2$ [MeV]
	 &28.04 &28.04 &27.15 &22.61 &17.54 \\ \hline\hline
    \end{tabular}
  \end{center}
\end{table}

  \subsection{Comparison with the Quark Cluster Model}
  \label{sec:Comparison_QCM}
It is interesting to compare our results with 
those of
the quark model~\cite{Takeuchi:2016ejt}.
Because of the color confinement,
the quark degrees of freedom affect only 
when
the 
relevant
hadrons come close to each other.
Investigating $q^4\overline{q}~(0s)^5$ states will
give a clue to the short-range part of 
the hadron interaction arising quark degrees of freedom.

The number of allowed states $q^4\overline{q}$  $(0s)^5$
is smaller than that of the meson-baryon states.
As shown in Table \ref{table:5qchannels}, 
the configuration of the isospin-1/2 three light quarks
is either color-singlet spin-1/2, color-octet spin-1/2, or color-octet spin-3/2.
Together with the spin-0 or -1 $c\overline{c}$ pair, 
there exist five spin-1/2, four spin-3/2, and one spin-5/2 $q^4\overline{q}$ $(0s)^5$ states.
The number of $S$-wave meson-baryon states is seven for $J=1/2$, five for $J=3/2$, 
and one for $J=5/2$. 
So, there are two [one] forbidden states
for the $J=1/2$ [$3/2$] system, 
where a certain combination of the meson-baryon states is forbidden
to exist as a $(0s)^5$ configuration.
The normalization of such states reduces to zero.
This leads to a strong repulsion to that particular combination of the meson-baryon states.
On the other hand,
there are channels where the normalization is larger than 1,
which brings the system an attraction.
The five quark states listed in Table~\ref{tbl2} have a normalization of 4/3.

Moreover, the color magnetic interaction (CMI) between quarks can contribute to
the hadron interaction.
In Ref.~\cite{Takeuchi:2016ejt}, 
the CMI, especially,
in the color-octet spin-3/2 configuration of three light quarks
brings to an attraction between $\bar{D}Y_c$.

It was reported in Ref.~\cite{Takeuchi:2016ejt} that
the quark cluster model gives a very shallow bound state for $J=5/2$ (4519.9 MeV),
a cusp and a resonance for $3/2$ (4379.3, 4457.8 MeV), and a resonance for $J=1/2$ channels
(4317.0 MeV).
Energy of each of the structures is close to the meson-baryon threshold,
and the widths of the resonances are 
as narrow as
a few MeV.

In the present work,
a bound state appears 
in
the $J^P=5/2^-$ channel when the strength of the short-range interaction
is about 
$f/f_0=25$ (Fig.~\ref{fig:results_52m}). 
We may consider that this strength roughly corresponds to that of the quark cluster model
because there is a shallow bound state in the channel.
Suppose the strength determined in the $J^P=5/2^-$ channel can also apply to the other channels, then
there are two resonances in the $J^P=3/2^-$ channels 
at around the same energies as those of the quark cluster model (Fig.~\ref{fig:results_32m_Sum}). 
In the $J^P=1/2^-$ channel, 
there are two resonances at  
$f/f_0=25$;
one of them corresponds to the quark model results, but additional resonance 
appears at around 
$\bar{D}^\ast\Sigma^{\ast}_{\rm c}$ 
threshold (Fig.~\ref{fig:results_12m_Sum}).
With this exception, the results of the present work are similar to the
quark model one. 
In the present approach, coupling to the five-quark states gives an attraction
to the meson-baryon channel,
which plays the same role as the ones from the above mentioned attraction in the quark model.

  \clearpage
  \subsection{Numerical results of the hidden-bottom sector}
  \label{sec:Numerical_results_bottom}
We discuss the hidden-bottom meson-baryon molecules in this section.
The basic features of the potentials are unchanged from those of the
hidden-charm, except that 
the cutoff parameters of the
OPEP are different as summarized in Sec.~\ref{sec:OPEP}.
However, 
the hadron masses in the bottom sector are larger than those in
the charm sector, and the mass splittings of the HQS multiplet ($B$ and
$B^\ast$, and $\Sigma_{\rm b}$ and $\Sigma^\ast_{\rm b}$) are
small.
Because of these facts, more states are expected for the bottom sector.
As a matter of fact, we find that only the OPEP provides sufficiently
strong attraction to generate several bound and resonant states.
The obtained energies only with the OPEP are summarized in Table.~\ref{table:energy_bottom_OPEP}.
Since the OPEP yields the strong attraction, 
we will see that both the OPEP and the $5q$ potentials have an important role to produce the energy spectra,
while the $S$-factor of the $5q$ potential designs the spectra in the hidden-charm sector.

  \begin{table}[t]
   \begin{center}
    \caption{\label{table:energy_bottom_OPEP}
    Energy spectra of the hidden-bottom molecules 
    only with the OPEP.
    The energy $E$ and the half decay width $\Gamma/2$
    are given in units of MeV.
    The lowest threshold $B\Lambda_{\rm b}$ is at 10898.51 MeV.
    }    
    \begin{tabular}{cc|ccc}
     \hline\hline
     $J^P=1/2^-$ &$E$ [MeV]
     &$10898$  &$10943$  &$11151$ \\     
     & $\Gamma/2$ [MeV] &--- &$1.80\times 10^{-2}$ &$2.01$ \\ \hline
     $J^P=3/2^-$ &$E$ [MeV] &$10942$ \\
     & $\Gamma/2$ [MeV] &$3.08\times 10^{-2}$ & \\
     \hline\hline
    \end{tabular}
   \end{center}
  \end{table}

\begin{figure}[t]
 \begin{center}
   \begin{tabular}{ccc}
    $J^P=1/2^-$&&\\    
    (i) $(S_{b\bar{b}},S_{3q})=(0,\frac{1}{2})$ 
    &	(ii) $(S_{b\bar{b}},S_{3q})=(1,\frac{1}{2})$ 
	&  (iii) $(S_{b\bar{b}},S_{3q})=(1,\frac{3}{2})$ \\ 
   \includegraphics[width=0.34\linewidth,clip]{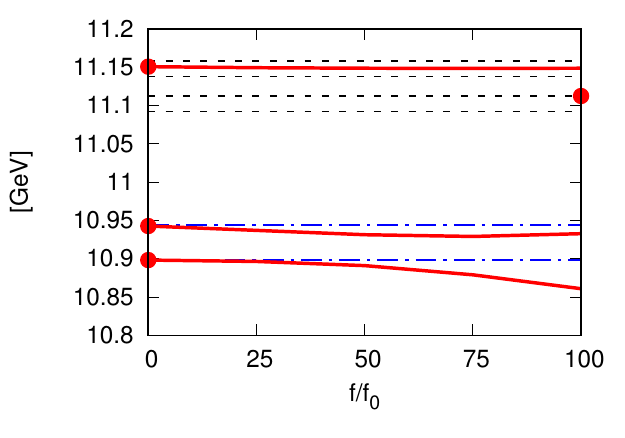}&
   \includegraphics[width=0.34\linewidth,clip]{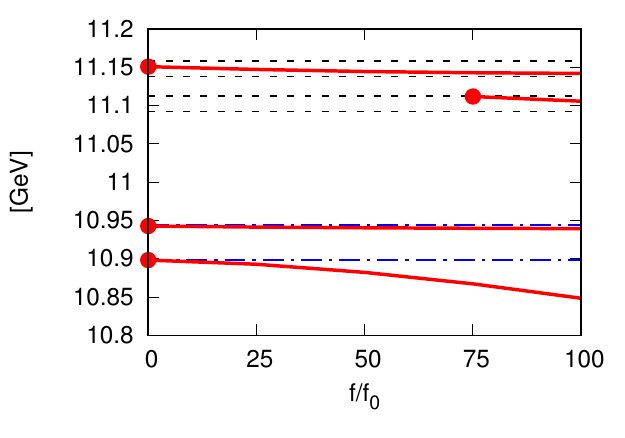}&
   \includegraphics[width=0.34\linewidth,clip]{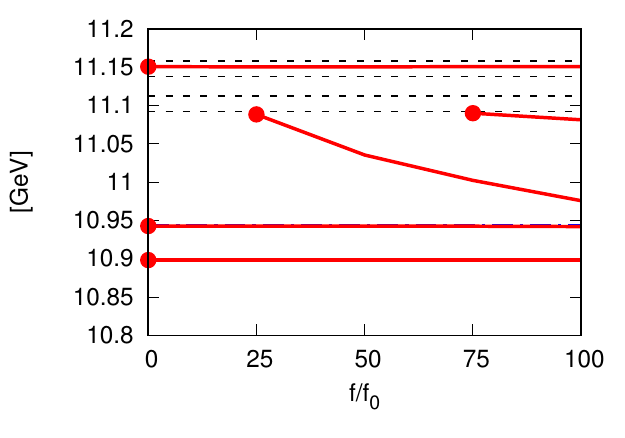}\\    
   \end{tabular}  
  \caption{\label{fig:results_12m_BYb} 
  (Color online) 
  Bound and resonant states of the hidden-bottom
  molecules with various coupling constants
  $f$ for $J^P=1/2^-$, using 
  the OPEP and one of the three $5q$ potentials derived from the configuration 
  (i) $(S_{b\bar{b}},S_{3q})=(0,1/2)$,
  (ii) $(1,1/2)$, 
  or 
  (iii) $(1,3/2)$.
  The horizontal axis shows the ratio $f/f_0$, where $f_0$ is the
  reference value defined in Sec.~\ref{sec:Model_parameters}.
  Solid line shows the obtained state. 
  Filled circle is the starting point where the states appear.
  Dashed lines are the $B\Sigma_{\rm b}$, $B\Sigma^{\ast}_{\rm b}$,
  $B^{\ast}\Sigma_{\rm b}$, and $B^{\ast}\Sigma^{\ast}_{\rm b}$
  thresholds.
  Dot-dashed lines are the $B\Lambda_{\rm b}$ and
  $B^{\ast}\Lambda_{\rm b}$ thresholds.
  }  
 \end{center}
\end{figure}

In Fig.~\ref{fig:results_12m_BYb} and Tables~\ref{table:energyspectra_12m_BYb_i}-\ref{table:energyspectra_12m_BYb_iii},
the strength $f$ dependence of the energy spectra obtained for
$J^P=1/2^-$ by using the OPEP and one of the three $5q$ potentials is shown.
The three $5q$ potentials are from the configurations 
(i) $(S_{b\bar{b}},S_{3q})=(0,1/2)$,
(ii) $(1,1/2)$, and (iii) $(1,3/2)$ which
are the same as discussed in the hidden-charm sector.
In Fig.~\ref{fig:results_12m_BYb} (i),
we find three states appearing for
$f/f_0\geq 0$ below
the three thresholds of $B\Lambda_{\rm b}$, $B^\ast\Lambda_{\rm b}$, and
$B^\ast\Sigma^\ast_{\rm b}$.
These states originate in those obtained only by using the OPEP in Table~\ref{table:energy_bottom_OPEP}.
As $f$ is increased, and reaches around $f/f_0\sim 100$,
another state appears below the $B\Sigma^\ast_{\rm b}$ threshold.
Here, 
we find that the $S$-factor of the $5q$ potential is zero in the $B\Sigma^\ast_{\rm b}$ component, while 
 the large $S$-factor is obtained in the $B^\ast\Lambda_{\rm b}$ and $B^\ast\Sigma^\ast_{\rm b}$ components.
 In producing the state, not only the $5q$ potential, but also the OPEP have the important role.

In Figs.~\ref{fig:results_12m_BYb} (ii) and (iii), and Tables~\ref{table:energyspectra_12m_BYb_ii} and 
\ref{table:energyspectra_12m_BYb_iii},
we show the energy spectra for using the $5q$ potentials from the other quark configurations (ii) and (iii).
These energy spectra also show the three states for $f/f_0\geq 0$ originating in those produced only by the OPEP.
In Fig.~\ref{fig:results_12m_BYb} (ii) , 
one resonance appears below the $B\Sigma^\ast_{\rm b}$,
as $f$ is increased.
In Fig.~\ref{fig:results_12m_BYb} (iii), 
two resonances 
appear 
below the $B\Sigma_{\rm b}$ threshold,
where
the large $S$-factor of the $5q$ potential is obtained in the $B\Sigma_{\rm b}$ component.

In Fig.~\ref{fig:results_12m_SUM_BYb} and Table~\ref{table:energyspectra_12m_BYb_SUM},
the results are shown with the full potential including OPEP and the sum
of the three $5q$ potentials for $J^P=1/2^-$.
The three states 
appearing below the $B\Lambda_{\rm b}$,
$B^\ast\Lambda_{\rm b}$ and $B^\ast\Sigma_{\rm b}$ thresholds for $f/f_0\geq 0$
originate those obtained only by using the OPEP.
Moreover, we obtain three resonances 
as $f$ is increased.

\begin{figure}[t]
 \begin{center}
   \begin{tabular}{ccc}
    $J^P=3/2^-$&&\\
    (i) $(S_{b\bar{b}},S_{3q})=(0,\frac{3}{2})$ 
    &(ii) $(S_{b\bar{b}},S_{3q})=(1,\frac{1}{2})$ 
	& (iii) $(S_{b\bar{b}},S_{3q})=(1,\frac{3}{2})$ \\ 
   \includegraphics[width=0.34\linewidth,clip]{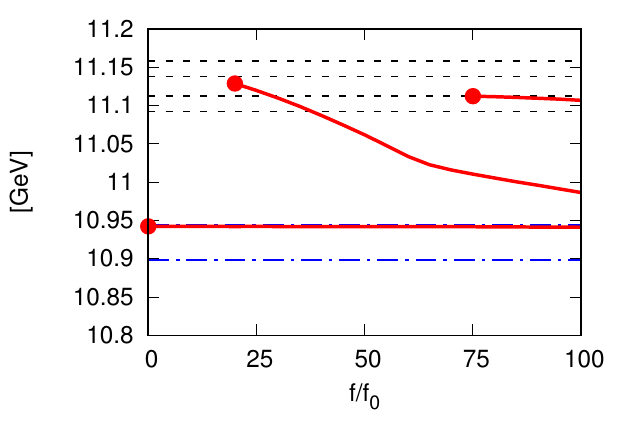}&
   \includegraphics[width=0.34\linewidth,clip]{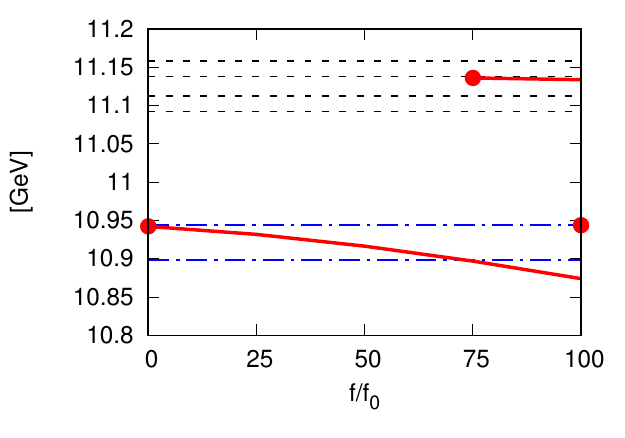}&
   \includegraphics[width=0.34\linewidth,clip]{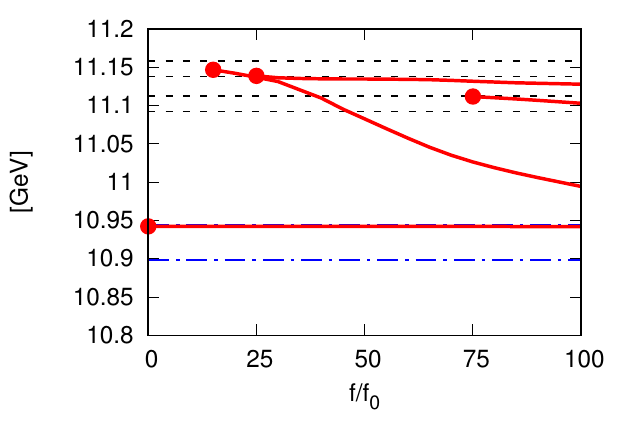}\\
   \end{tabular}  
  \caption{\label{fig:results_32m_BYb}
  (Color online)
  The same as Fig.~\ref{fig:results_12m_BYb} for the bound and resonant
  states of the hidden-bottom molecules for $J^P=3/2^-$ using the
  OPEP and 
  one of the three $5q$ potentials derived from the configuration 
  (i)~$(S_{b\bar{b}},S_{3q})=(0,3/2)$,
  (ii)~$(1,1/2)$, or (iii)~$(1,3/2)$.
  }
 \end{center}
\end{figure}

\begin{figure}[t]
 \begin{center}
  $J^P=3/2^-$\\
  \includegraphics[width=0.6\linewidth,clip]{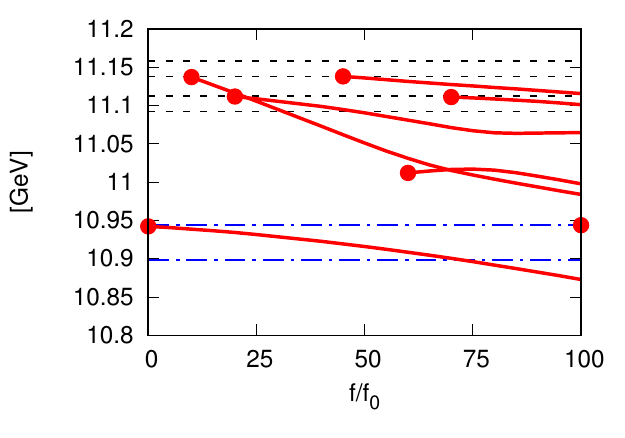}
  \caption{\label{fig:results_32m_SUM_BYb}
  (Color online)
  The same as Fig.~\ref{fig:results_12m_BYb} for the bound and resonant
  states of the hidden-bottom molecules for $J^P=3/2^-$ using the
  OPEP and the sum of the three $5q$ potentials.
  }  
 \end{center}
\end{figure}

The states are also found in $J^P=3/2^-$.
Fig.~\ref{fig:results_32m_BYb} and Tables~\ref{table:energyspectra_32m_BYb_i}-\ref{table:energyspectra_32m_BYb_iii}
show the results with the OPEP and
one of the $5q$ potentials derived from the quark configurations
(i) $(S_{b\bar{b}},S_{3q})=(0,3/2)$, (ii) $(1,1/2)$, and (iii) $(1,3/2)$.
In Figs. \ref{fig:results_32m_BYb} (i), (ii), and (iii),
one state appears below the $B^\ast \Lambda_{\rm b}$ threshold for $f/f_0\geq 0$,
which originates in the state obtained only by using the OPEP in Table~\ref{table:energy_bottom_OPEP}.
In addition, we obtain the states as $f$ is increased.
In Fig.~\ref{fig:results_32m_BYb} (i),
two resonances appear below the $B\Sigma^\ast_{\rm b}$ and 
$B^\ast\Sigma_{\rm b}$ thresholds, where the large $S$-factors of the $5q$ potential 
are obtained in the $B\Sigma^\ast_{\rm b}$, $B^\ast\Sigma_{\rm b}$, and $B^\ast\Sigma^\ast_{\rm b}$ components.
In Fig.~\ref{fig:results_32m_BYb} (ii), 
two resonances appear below the $B^\ast \Lambda_{\rm b}$ and 
$B^\ast\Sigma_{\rm b}$ thresholds, where the large $S$-factor is 
obtained in the $B^\ast\Lambda_{\rm b}$ component.
In Fig.~\ref{fig:results_32m_BYb} (ii), 
three resonances appear near the $B\Sigma^\ast_{\rm b}$,
$B^\ast\Sigma_{\rm b}$, and $B^\ast\Sigma^\ast_{\rm b}$ thresholds, where the large $S$-factors  
are obtained in the $B\Sigma^\ast_{\rm b}$ and $B^\ast\Lambda_{\rm b}$ components.
In the results obtained for $J^P=3/2^-$,
several spectra can be explained by the large $S$-factors of the $5q$ potential,
while both the OPEP and $5q$ potential are important in producing the other states.
The energy spectra with the full potential including the OPEP and the sum of the three $5q$ potentials for
$J^P=3/2^-$ are displayed in Fig.~\ref{fig:results_32m_SUM_BYb} 
and Tables~\ref{table:energyspectra_32m_BYb_SUM}-\ref{table:energyspectra_32m_BYb_SUM_2}.
The state below the $B^\ast\Lambda_{\rm b}$ threshold for $f/f_0\geq 0$
originates the state obtained only by using the OPEP.
Moreover, many states appear, when the $5q$ potential is switched on.

\begin{figure}[t]
  \begin{center}
    \begin{tabular}{c}
     $J^P=5/2^-$ \\  
     \includegraphics[width=0.6\linewidth,clip]{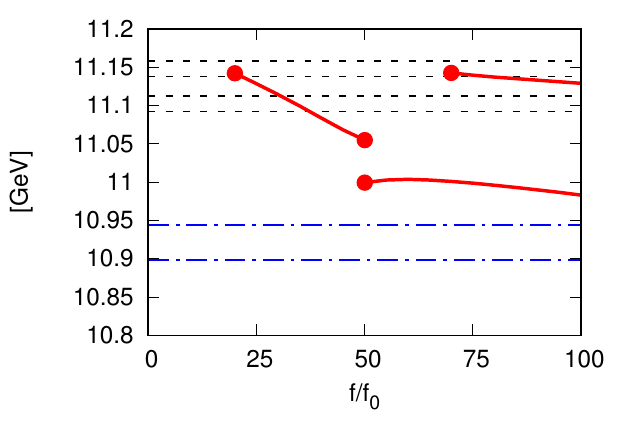}   \\
    \end{tabular}
  \caption{\label{fig:results_52m_BYb}
   (Color online)
   The same as Fig.~\ref{fig:results_12m_BYb} for the resonant
   states of the hidden-bottom molecules for $J^P=5/2^-$ using the
   OPEP and the $5q$ potential from the 
   configuration $(S_{b\bar{b}},S_{3q})=(1,\frac{3}{2})$.
  }  
  \end{center}
\end{figure}

Figure~\ref{fig:results_52m_BYb} 
and Table~\ref{table:energyspectra_52m_BYb} 
give the strength $f$ dependence of the
energy spectra for $J^P=5/2^-$ with
the OPEP and the $5q$ potential from the quark configuration $(S_{b\bar{b}},S_{3q})=(1,3/2)$.
For $J^P=5/2^-$, we do not obtain any state when only the OPEP is employed.
The three resonances are obtained, as $f$ of the $5q$ potential is increased.
Two resonances appear near the $B^\ast\Sigma_{\rm b}$ threshold.
The state obtained for $20\leq f/f_0 \leq 50$ disappears as $f$ is increased, 
whose width becomes large.
Moreover, one resonance appears above the $B^\ast\Lambda_{\rm b}$ threshold for $f/f_0\geq 50$.

In the hidden-bottom sector, 
the OPEP is strong enough to produce 
states
due to 
the mixing effect enhanced by the small mass splitting between $B$ and $B^\ast$, 
and $\Sigma_{\rm b}$ and $\Sigma^\ast_{\rm b}$.
Thus, both the OPEP and the $5q$ potential play the important role to produce many states,
while the $5q$ potential has the dominant role to yield the states in the hidden-charm sector.
Since the attraction from the OPEP is enhanced and 
the kinetic term is suppressed due to the large hadron masses,
the hidden-bottom pentaquarks are more likely to form rather than the hidden-charm pentaquarks.

\begin{figure}[t]
 \begin{center}
  $J^P=1/2^-$\\
  \includegraphics[width=0.6\linewidth,clip]{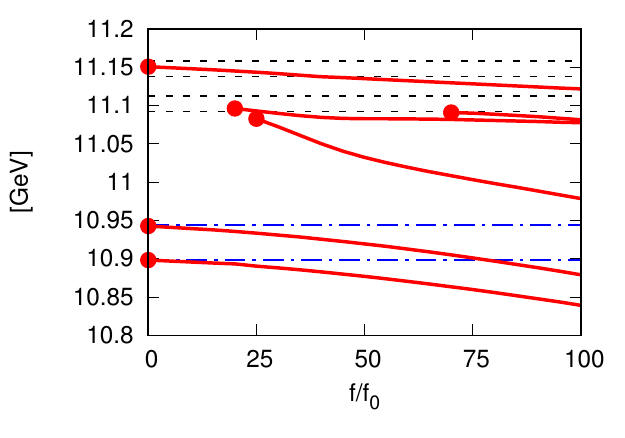}
  \caption{\label{fig:results_12m_SUM_BYb}
  The same as Fig.~\ref{fig:results_12m_BYb} for the bound and resonant
  states of the hidden-bottom molecules for $J^P=1/2^-$ using the
  OPEP and the sum of the three $5q$ potentials.
  }
  \label{fig:results_12m_SUM_BYb}
 \end{center}
\end{figure}


\begin{table}[t]
 \caption{\label{table:energyspectra_12m_BYb_i}
 Energy spectra of the hidden-bottom
 molecules for $J^P=1/2^-$
 using the OPEP and the $5q$ potential from 
 the configuration (i) $(S_{b\bar{b}},S_{3q})=(0,1/2)$.
 The energy $E$ and half decay width $\Gamma/2$ in the various coupling constants $f/f_0$ are shown.
 The third row is for the point, where the state appears.
 The fourth, fifth, sixth and seventh rows show the obtained values
 with $f=25f_0$, $50f_0$, $75f_0$, and $100f_0$, respectively.
 The values are given in units of MeV.
 The lowest threshold $B\Lambda_{\rm b}$ is at 10898.51 MeV,
 and the state whose energy is lower than the threshold is a bound state.
 } 
  \begin{center}
    \begin{tabular}{cc||c|cccc}
     \hline\hline
    (i) $(0,1/2)$& $f/f_0$
     &0 &25 &50 &75 &100 \\
     &$E$ [MeV]
	 &11151 &11150 &11149 &11149 &11149 \\ 
     &$\Gamma/2$ [MeV]
	 &2.01 &3.05 &4.25 &5.32 &6.08 \\ \hline
     & $f/f_0$
     &100 &25 &50 &75 &100 \\
     &$E$ [MeV]
	 &11113 &--- &--- &--- &11113 \\ 
     &$\Gamma/2$ [MeV]
	 &6.43 &--- &--- &--- &6.43 \\ \hline
     & $f/f_0$
     &0 &25 &50 &75 &100 \\
     &$E$ [MeV]
	 &10943 &10937 &10932 &10929 &10933 \\ 
     &$\Gamma/2$ [MeV]
	 &1.80$\times 10^{-2}$ &0.55 &2.92 &7.13 &7.89 \\ \hline     
     & $f/f_0$
     &0 &25 &50 &75 &100 \\
     &$E$ [MeV]
	 &10898 &10897 &10891 &10879 &10861 \\ 
     &$\Gamma/2$ [MeV]
	 &--- &--- &--- &--- &--- \\ 
     \hline\hline
    \end{tabular}
  \end{center}
\end{table}

\begin{table}[t]
 \caption{\label{table:energyspectra_12m_BYb_ii}
The same as Table~\ref{table:energyspectra_12m_BYb_i} for the
 energy spectra of the
 hidden-bottom molecules for
 $J^P=1/2^-$ using the OPEP and the $5q$ potential from the configuration 
 (ii) $(S_{b\bar{b}},S_{3q})=(1,1/2)$.
 } 
  \begin{center}
    \begin{tabular}{cc||c|cccc}
     \hline\hline
     (ii) $(1,1/2)$ & $f/f_0$
     &0 &25 &50 &75 &100 \\
     &$E$ [MeV]
	 &11151 &11147 &11145 &11143 &11142 \\ 
     &$\Gamma/2$ [MeV]
	 &2.01 &1.75 &2.76 &4.22 &5.52 \\ \hline
     & $f/f_0$
     &75 &25 &50 &75 &100  \\
     &$E$ [MeV]
	&11112 &--- &--- &11112 &11106 \\ 
     &$\Gamma/2$ [MeV]
	 &7.68 &--- &--- &7.68 &5.25  \\ \hline
     & $f/f_0$
     &0 &25 &50 &75 &100 \\
     &$E$ [MeV]
	&10943 &10941 &10941 &10940 &10939 \\ 
     &$\Gamma/2$ [MeV]
	 &1.80$\times 10^{-2}$ &0.19 &0.31 &0.33 &0.22 \\ \hline
     & $f/f_0$
     &0 &25 &50 &75 &100 \\
     &$E$ [MeV]
	 &10898 &10893 &10882 &10867 &10848 \\ 
     &$\Gamma/2$ [MeV]
	 &--- &--- &--- &--- &--- \\ \hline     
    \end{tabular}
  \end{center}
\end{table}

\begin{table}[t]
 \caption{\label{table:energyspectra_12m_BYb_iii}
The same as Table~\ref{table:energyspectra_12m_BYb_i} for the
 energy spectra of the
 hidden-bottom molecules for
 $J^P=1/2^-$ using the OPEP and the $5q$ potential from the configuration 
(iii) $(S_{b\bar{b}},S_{3q})=(1,3/2)$.
 } 
  \begin{center}
    \begin{tabular}{cc||c|cccc}
     \hline\hline
     (iii) $(1,3/2)$
     & $f/f_0$
     &0 &25 &50 &75 &100 \\
     &$E$ [MeV]
	&11151 &11151 &11151 &11151 &11151 \\ 
     &$\Gamma/2$ [MeV]
	 &2.01 &2.63 &2.89 &2.92 &2.91 \\ \hline
     & $f/f_0$
     &75 &25 &50 &75 &100 \\
     &$E$ [MeV]
	&11090 &--- &--- &11090 &11082 \\  
     &$\Gamma/2$ [MeV]
	 &0.37 &--- &--- &0.37 &0.30 \\ \hline
     & $f/f_0$
     &25  &25 &50 &75 &100 \\
     &$E$ [MeV]
	&11089 &11089 &11036 &11002 &10976 \\  
     &$\Gamma/2$ [MeV]
	 &29.54 &29.54 &26.93 &12.38 &4.35 \\ \hline
     & $f/f_0$
     &0 &25 &50 &75 &100 \\
     &$E$ [MeV]
	&10943 &10943 &10943 &10943 &10942 \\  
     &$\Gamma/2$ [MeV]
	 &1.80$\times 10^{-2}$ &0.13 &0.13 &0.13 &0.17 \\ \hline
     & $f/f_0$
     &0 &25 &50 &75 &100 \\
     &$E$ [MeV]
	 &10898 &10898 &10898 &10898 &10898 \\ 
     &$\Gamma/2$ [MeV]
	 &--- &--- &--- &--- &--- \\
     \hline\hline
    \end{tabular}
  \end{center}
\end{table}

\begin{table}[t]
 \caption{\label{table:energyspectra_12m_BYb_SUM}
 The same as Table~\ref{table:energyspectra_12m_BYb_i} for the
 energy spectra of the hidden-bottom molecules for
 $J^P=1/2^-$ using the OPEP and  the sum of the three $5q$ potentials.
 }
  \begin{center}
    \begin{tabular}{cc||c|cccc}
     \hline\hline
     SUM
     & $f/f_0$
     &0 &25 &50 &75 &100 \\
     &$E$ [MeV]
	&11151 &11144 &11135 &11129 &11122 \\  
     &$\Gamma/2$ [MeV]
	 &2.01 &2.67 &0.60 &0.58 &0.60 \\ \hline
     & $f/f_0$
     &70 &25 &50 &75 &100 \\
     &$E$ [MeV]
	&11091 &--- &--- &11090 &11082 \\  
     &$\Gamma/2$ [MeV]
	 &0.36 &--- &--- &0.44 &0.75 \\ \hline
     & $f/f_0$
     &20 &25 &50 &75 &100 \\
     &$E$ [MeV]
	&11096 &11093 &11083 &11081 &11078 \\  
     &$\Gamma/2$ [MeV]
	 &44.69 &11.35 &14.15 &31.45 &39.32 \\ \hline
     & $f/f_0$
     &25 &25 &50 &75 &100 \\
     &$E$ [MeV]
	&11083 &11083 &11033 &11003 &10979 \\  
     &$\Gamma/2$ [MeV]
	 &78.77 &78.77 &40.76 &14.49 &4.03 \\ \hline
     & $f/f_0$
     &0 &25 &50 &75 &100 \\
     &$E$ [MeV]
	&10943 &10934 &10920 &10901 &10879 \\ 
     &$\Gamma/2$ [MeV]
	 &1.80$\times 10^{-2}$ &1.91$\times 10^{-2}$ &5.80$\times 10^{-2}$ &0.12 &--- \\ \hline
     & $f/f_0$
     &0 &25 &50 &75 &100 \\
     &$E$ [MeV]
	&10898 &10891 &10877 &10860 &10839 \\  
     &$\Gamma/2$ [MeV]
	 &--- &--- &--- &--- &--- \\
     \hline\hline     
    \end{tabular}
  \end{center}
\end{table}

\begin{table}[t]
 \caption{\label{table:energyspectra_32m_BYb_i}
The same as Table~\ref{table:energyspectra_12m_BYb_i} for the
 energy spectra of the
 hidden-bottom molecules for
 $J^P=3/2^-$ using the OPEP and the $5q$ potential from the configuration 
(i) $(S_{b\bar{b}},S_{3q})=(0,3/2)$. 
 } 
  \begin{center}
    \begin{tabular}{cc||c|cccc}
     \hline\hline
     (i) $(0,3/2)$
     & $f/f_0$
     &75 &25 &50 &75 &100 \\
     &$E$ [MeV]
	 &11112 &--- &--- &11112 &11107 \\ 
     &$\Gamma/2$ [MeV]
	 &1.13 &--- &--- &1.13 &1.13 \\ \hline
     & $f/f_0$
     &20 &25 &50 &75 &100 \\
     &$E$ [MeV]
	&11129 &11120 &11062 &11011 &10987 \\  
     &$\Gamma/2$ [MeV]
	 &57.15 &59.69 &64.94 &34.53 &16.76 \\ \hline
     & $f/f_0$
     &0 &25 &50 &75 &100 \\
     &$E$ [MeV]
	&10942 &10942 &10942 &10942 &10941 \\  
     &$\Gamma/2$ [MeV]
	 &3.08$\times 10^{-2}$ &0.15 &0.17 &0.16 &0.23 \\
     \hline\hline
    \end{tabular}
  \end{center}
\end{table}

\begin{table}[t]
 \caption{\label{table:energyspectra_32m_BYb_ii}
 The same as Table~\ref{table:energyspectra_12m_BYb_i} for the
 energy spectra of the
 hidden-bottom molecules for
 $J^P=3/2^-$ using the OPEP and the $5q$ potential from the configuration 
 (ii) $(S_{b\bar{b}},S_{3q})=(1,1/2)$.
 } 
  \begin{center}
    \begin{tabular}{cc||c|cccc}
     \hline\hline
     (ii) $(1,1/2)$
     & $f/f_0$
     &75 &25 &50 &75 &100 \\
     &$E$ [MeV]
	 &11136 &--- &--- &11136 &11134 \\ 
     &$\Gamma/2$ [MeV]
	 &19.45 &--- &--- &19.45 &11.86 \\ \hline
     & $f/f_0$
     &100 &25 &50 &75 &100 \\
     &$E$ [MeV]
	&10944 &--- &--- &--- &10944 \\  
     &$\Gamma/2$ [MeV]
	 &0.11 &--- &--- &--- &0.11 \\ \hline
     & $f/f_0$
     &0 &25 &50 &75 &100 \\
     &$E$ [MeV]
	&10942 &10932 &10917 &10897 &10874 \\  
     &$\Gamma/2$ [MeV]
	 &3.08$\times 10^{-2}$ &0.13 &0.11 &--- &--- \\ 
     \hline\hline
    \end{tabular}
  \end{center}
\end{table}

\begin{table}[t]
 \caption{\label{table:energyspectra_32m_BYb_iii}
 The same as Table~\ref{table:energyspectra_12m_BYb_i} for the
 energy spectra of the
 hidden-bottom molecules for
 $J^P=3/2^-$ using the OPEP and the $5q$ potential from the configuration 
 (iii) $(S_{b\bar{b}},S_{3q})=(1,3/2)$.
 }
  \begin{center}
    \begin{tabular}{cc||c|cccc}
     \hline\hline
     (iii) $(1,3/2)$
     & $f/f_0$
     &25 &25 &50 &75 &100 \\
     &$E$ [MeV]
	&11139 &11139 &11135 &11132 &11128 \\ 
     &$\Gamma/2$ [MeV]
	 &22.58 &22.58 &16.00 &11.53 &12.61 \\ \hline
     & $f/f_0$
     &75 &25 &50 &75 &100 \\
     &$E$ [MeV]
	&11112 &--- &--- &11112 &11103 \\ 
     &$\Gamma/2$ [MeV]
	 &1.91 &--- &--- &1.91 &1.15 \\ \hline
     & $f/f_0$
     &15 &25 &50 &75 &100 \\
     &$E$ [MeV]
	&11147 &11137 &11083 &11027 &10995 \\ 
     &$\Gamma/2$ [MeV]
	 &47.21 &45.51 &40.07 &28.14 &11.19 \\ \hline
     & $f/f_0$
     &0 &25 &50 &75 &100 \\
     &$E$ [MeV]
	&10942 &10942 &10942 &10942 &10942 \\ 
     &$\Gamma/2$ [MeV]
	 &3.08$\times 10^{-2}$ &8.92$\times 10^{-3}$ &1.01$\times 10^{-2}$ &1.21$\times 10^{-2}$ &1.68$\times 10^{-2}$ \\ 
     \hline\hline     
    \end{tabular}
  \end{center}
\end{table}

\begin{table}[t]
   \begin{center}
     \caption{\label{table:energyspectra_32m_BYb_SUM}
    The same as Table~\ref{table:energyspectra_12m_BYb_i} for the
    energy spectra of the hidden-bottom molecules for
    $J^P=3/2^-$ using the OPEP and  the sum of the three $5q$ potentials.
    } 
    \begin{tabular}{cc||c|cccc}
     \hline\hline
     SUM
     & $f/f_0$
     &45 &25 &50 &75 &100 \\
     &$E$ [MeV]
	&11138 &--- &11136 &11126 &11116 \\ 
     &$\Gamma/2$ [MeV]
	 &5.13 &--- &5.71 &3.78 &1.94 \\ \hline
     & $f/f_0$
     &70 &25 &50 &75 &100 \\
     &$E$ [MeV]
	&11111 &--- &--- &11110 &11101 \\ 
     &$\Gamma/2$ [MeV]
	 &0.27 &--- &--- &0.35 &0.70 \\ \hline
     & $f/f_0$
     &20 &25 &50 &75 &100 \\
     &$E$ [MeV]
	&11112 &11109 &11091 &11067 &11065 \\ 
     &$\Gamma/2$ [MeV]
	 &4.40 &5.57 &11.82 &28.88 &51.60 \\ \hline
     & $f/f_0$
     &60 &25 &50 &75 &100 \\
     &$E$ [MeV]
	&11012 &--- &--- &11017 &10998 \\ 
     &$\Gamma/2$ [MeV]
	 &53.76 &--- &--- &37.95 &10.85 \\ 
     \hline\hline
    \end{tabular}
   \end{center}
\end{table}

\begin{table}[t]
   \begin{center}
     \caption{\label{table:energyspectra_32m_BYb_SUM_2}
    Continued from Table~\ref{table:energyspectra_32m_BYb_SUM}.
    }
    \begin{tabular}{cc||c|cccc}
     \hline\hline
    SUM & $f/f_0$
     &10 &25 &50 &75 &100 \\
     &$E$ [MeV]
	&11137 &11106 &11051 &11010 &10984 \\ 
     &$\Gamma/2$ [MeV]
	 &52.77 &58.70 &54.22 &29.71 &12.94 \\ \hline
     & $f/f_0$
     &100 &25 &50 &75 &100 \\
     &$E$ [MeV]
	&10944 &--- &--- &--- &10944 \\ 
     &$\Gamma/2$ [MeV]
	 &4.70$\times 10^{-3}$ &--- &--- &--- &4.70$\times 10^{-3}$ \\ \hline
     & $f/f_0$
     &0 &25 &50 &75 &100 \\
     &$E$ [MeV]
	&10942 &10932 &10916 &10896 &10873 \\ 
     &$\Gamma/2$ [MeV]
	 &3.08$\times 10^{-2}$ &7.83$\times 10^{-3}$ &1.97$\times 10^{-3}$ &--- &--- \\
     \hline\hline
    \end{tabular}
   \end{center}
\end{table}

\begin{table}[t]
 \caption{\label{table:energyspectra_52m_BYb}
 The same as Table~\ref{table:energyspectra_12m_BYb_i} for the
 energy spectra of the hidden-bottom molecules for
 $J^P=5/2^-$ using the OPEP and the $5q$ potential from the configuration $(S_{b\bar{b}},S_{3q})=(1,3/2)$.
 } 
  \begin{center}
    \begin{tabular}{cc||c|cccc}
     \hline\hline
     $(1,3/2)$
     & $f/f_0$
     &70 &25 &50 &75 &100 \\
     &$E$ [MeV]
	 &11142.84 &--- &--- &11139.85 &11129.35 \\
     &$\Gamma/2$ [MeV]
	 &15.89 &--- &--- &12.66 &5.15 \\ \hline
     & $f/f_0$
     &20 &25 &50 &75 &100 \\
     &$E$ [MeV]
	 &11142.42 &11128.79 &11055.16 &--- &--- \\
     &$\Gamma/2$ [MeV]
	 &123.11 &125.94 &153.98 &--- &--- \\ \hline
     & $f/f_0$
     &50 &25 &50 &75 &100 \\
     &$E$ [MeV]
	 &10999.46 &--- &10999.46 &10998.89 &10983.33 \\
     &$\Gamma/2$ [MeV]
	 &71.82 &--- &71.82 &36.75 &17.97 \\
     \hline\hline
    \end{tabular}
  \end{center}
\end{table}

  \section{Summary}
  \label{sec:summary}
In this paper, we have studied hidden-charm and hidden-bottom pentaquark states.  
Since the observed $P_c$'s are in the open-charm threshold region,
we have performed a coupled channel analyses with various meson-baryon
states which may generate bound and resonant states.
In such an analysis, the hadronic interaction is the most important input.    
At long distances, we employ the one-pion exchange potential which is
best known among various hadron interactions.
As discussed and emphasized in many works, the OPEP provides attraction
when the tensor force is at work through the $SD$ coupled channels.
This is crucially important for the formation of the exotic pentaquark states.  

Contrary, for short range interaction which is far less known, 
we inferred from a recent quark cluster model analysis pointing out the
importance of the colorful $5q$ configurations.
We have included these $5q$ configurations in the coupled channel problems as one-particle states. 
By eliminating them we have derived an effective interaction at short distances.  
Since all the expected $5q$ states locate above the meson-baryon threshold region,
the resulting effective interaction is attractive, which can be another
driving force for the generation of the pentaquark states.
The coupling of this interaction to various meson-baryon channels is estimated by the spectroscopic factor.
Therefore, our model contains essentially only one parameter which is
the overall strength of the short range interaction $f$.
Then results are shown for various $f$ up to the maximum strength which
we expect from our current knowledge of the hadron interaction.

For the charm sector, when the $5q$ interaction is turned on, bound and
resonant states are generated for various spins, $1/2^-$, $3/2^-$, and $5/2^-$.
Among them, $3/2^-$ state with mass around 4460 MeV and width around 25
MeV
(see Table~\ref{table:energyspectra_32m}) is a candidate of the observed $P_c$,
though the spin parity identification is not the suggested one.  
Therefore, in this paper,
we have further concentrated on the mechanism how the pentaquark states are generated.   

For the bottom sector, due to the suppression of the kinetic energy, we
have seen abundant pentaquark states even only by the OPEP.
These are the rather robust predictions of our analysis.  
Therefore, with possible further attractions from the short range
interaction,
we indeed expect many exotic pentaquark states.  
In this way, we suggest experimental analysis to search for further states in the bottom region.  

We have also compared our present analysis with the previous quark cluster model one.  
We have found similarities between them, and therefore,
our approach provides a good method to make physical interpretations for
the results of the quark cluster model.  

In the present analysis we have studied negative parity states dominated by the $S$-wave configurations of open charm channels.
For more complete analysis, it is needed to include hidden-charm channels such as $J/\psi p$. 
In the case of the $Z_c(3900)$,
the importance of the mixing of $\bar{D}D^\ast-J/\psi \pi$ has been indicated by a lattice QCD simulation~\cite{Ikeda:2016zwx}.
It is also interesting to study positive parity states.
For this, we need $P$-wave excitations for both meson-baryon and for 5q states.
Moreover, couplings to such as $\bar{D}\Lambda_{\rm c}(2595)$ channel can be important 
because of their very close threshold to the $\bar{D}\Lambda_{\rm c}(2595)$ threshold, and to the reported $P_{\rm c}(4450)$ state~\cite{Burns:2015dwa}.
As discussed in Ref.~\cite{Geng:2017hxc}, such a coupling may show up a unique feature of the universal phenomena 
caused by the almost on-shell pion decaying from the $\Lambda_c(2595)$.
All these issues may be studied as interesting future investigations.

  \begin{acknowledgments}
   This work is supported by JSPS KAKENHI [the Grant-in-Aid for
   Scientific Research from Japan Society for the Promotion of Science
   (JSPS)] with Grant Nos.  JP16K05361 (S.T. and M.T.), JP17K05441(C) (A.H.), and JP26400273(C) (A.H.),
   by the Istituto Nazionale di Fisica Nucleare (INFN) Fellowship Programme (Y.Y), and by the Special Postdoctoral
   Researcher (SPDR) Program of RIKEN (Y.Y.).
  \end{acknowledgments}

  \appendix

  \section{Explicit form of the one-pion exchange potential}
  \label{sec:appendix_OPEP}
 The OPEP is given by the effective Lagrangians  in
 Eqs.~\eqref{eq:LagrangianDDpi} and \eqref{eq:LagrangianBBpi}.
 We use the static approximation where the energy transfer is neglected
 as compared to the momentum transfer.
 The OPEP for isospon $I=1/2$ is obtained by
\begin{align}
 &V^\pi_{\bar{D}^\ast\Sigma_{\rm c}-\bar{D}\Lambda_{\rm c}}(r)
 =-\frac{gg_4} {3\sqrt{2}f^2_\pi} 
 \left[\vec{\varepsilon}\,^\dagger\cdot\vec{\sigma} C(r)+S_{\varepsilon{\sigma}}(\hat{r})T(r)\right],\\
 &V^\pi_{\bar{D}^\ast\Sigma^\ast_{\rm c}-\bar{D}\Lambda_{\rm c}}(r)
 =\frac{gg_4} {\sqrt{6}f^2_\pi} 
 \left[\vec{\varepsilon}\,^\dagger\cdot\vec{\bar{\Sigma}} C(r)+S_{\varepsilon\bar{\Sigma}}(\hat{r})T(r)\right],\\
 &V^\pi_{\bar{D}\Sigma_{\rm c}-\bar{D}^\ast\Lambda_{\rm c}}(r)
 =-\frac{gg_4} {3\sqrt{2}f^2_\pi} 
 \left[\vec{\varepsilon}\cdot\vec{\sigma} C(r)+S_{\varepsilon{\sigma}}(\hat{r})T(r)\right],\\
 &V^\pi_{\bar{D}\Sigma^\ast_{\rm c}-\bar{D}^\ast\Lambda_{\rm c}}(r)
 =\frac{gg_4} {\sqrt{6}f^2_\pi} 
 \left[\vec{\varepsilon}\cdot\vec{\bar{\Sigma}} C(r)+S_{\varepsilon\bar{\Sigma}}(\hat{r})T(r)\right],\\
 &V^\pi_{\bar{D}^\ast\Sigma_{\rm c}-\bar{D}^\ast\Lambda_{\rm c}}(r)
 =-\frac{gg_4} {3\sqrt{2}f^2_\pi}  
 \left[\vec{S}\cdot\vec{\sigma} C(r)+S_{S{\sigma}}(\hat{r})T(r)\right],\\
 &V^\pi_{\bar{D}^\ast\Sigma^\ast_{\rm c}-\bar{D}^\ast\Lambda_{\rm c}}(r)
 =\frac{gg_4} {\sqrt{6}f^2_\pi}  
 \left[\vec{S}\cdot\vec{\bar{\Sigma}}^\dagger C(r)+S_{S\bar{\Sigma}}(\hat{r})T(r)\right],\\
 &V^\pi_{\bar{D}^\ast\Sigma_{\rm c}-\bar{D}\Sigma_{\rm c}}(r)
 = \frac{gg_1}{3f^2_\pi} 
 \left[\vec{\varepsilon}\,^\dagger\cdot\vec{\sigma} C(r)+S_{\varepsilon{\sigma}}(\hat{r})T(r)\right],\\
 &V^\pi_{\bar{D}^\ast\Sigma^{\ast}_{\rm c}-\bar{D}\Sigma_{\rm c}}(r)
 = \frac{gg_1}{2\sqrt{3}f^2_\pi} 
 \left[\vec{\varepsilon}\,^\dagger\cdot\vec{\bar{\Sigma}}\,^\dagger C(r)+S_{\varepsilon\bar{\Sigma}}(\hat{r})T(r)\right],\\
 &V^\pi_{\bar{D}^\ast\Sigma_{\rm c}-\bar{D}\Sigma^\ast_{\rm c}}(r)
 = \frac{gg_1}{2\sqrt{3}f^2_\pi} 
 \left[\vec{\varepsilon}\,^\dagger\cdot\vec{\bar{\Sigma}} C(r)+S_{\varepsilon\bar{\Sigma}}(\hat{r})T(r)\right],\\
 &V^\pi_{\bar{D}^\ast\Sigma^{\ast}_{\rm c}-\bar{D}\Sigma^\ast_{\rm c}}(r)
 = \frac{gg_1}{3f^2_\pi} 
 \left[\vec{\varepsilon}\,^\dagger\cdot\vec{\Sigma} C(r)+S_{\varepsilon\Sigma}(\hat{r})T(r)\right],\\
 &V^\pi_{\bar{D}^\ast\Sigma_{\rm c}-\bar{D}^\ast\Sigma_{\rm c}}(r)
 = -\frac{gg_1}{3f^2_\pi}  
 \left[\vec{S}\cdot\vec{\sigma} C(r)+S_{S\sigma}(\hat{r})T(r)\right],
 \label{eq:OPEP-P*B6-P*B6} \\
 &V^\pi_{\bar{D}^\ast\Sigma^\ast_{\rm c}-\bar{D}^\ast\Sigma_{\rm c}}(r)
 = \frac{gg_1}{2\sqrt{3}f^2_\pi} 
 \left[\vec{S}\cdot\vec{\bar{\Sigma}}^\dagger C(r)+S_{S\bar{\Sigma}}(\hat{r})T(r)\right],\\
 &V^\pi_{\bar{D}^\ast\Sigma^\ast_{\rm c}-\bar{D}^\ast\Sigma^\ast_{\rm c}}(r)
 = \frac{gg_1}{3f^2_\pi}  
 \left[\vec{S}\cdot\vec{\Sigma}C(r)+S_{S\Sigma}(\hat{r})T(r)\right].
\end{align}
The tensor operator
$S_{{\cal O}_{\bar{D}}{\cal O}_{Y_c}}(\hat{r})$ is 
defined by $S_{{\cal O}_{\bar{D}}{\cal O}_{Y_c}}(\hat{r})=3\vec{\cal
O}_{\bar{D}}\cdot\hat{r}\vec{\cal O}_{Y_c}\cdot\hat{r}-\vec{\cal O}_{\bar{D}}\cdot\vec{\cal O}_{Y_c}$ 
with the spin operators ${\cal O}_{\bar{D}}=\varepsilon, S$ for the
meson vertex and ${\cal O}_{Y_c}=\sigma, \bar{\Sigma}, \Sigma$ for the baryon vertex.
The polarization vector is defined by
$\vec{\varepsilon}\,^{(\pm)}=(\mp 1/\sqrt{2}, \pm i/\sqrt{2}, 0)$ and 
$\vec{\varepsilon}\,^{(0)}=(0, 0, 1)$.
The spin-one operator is
$\vec{S}=i\vec{\varepsilon}\times\vec{\varepsilon}\,^\dagger$,
$\vec{\sigma}$ is the Pauli matrices,
$\bar{\Sigma}^\mu$ is given by
\begin{align}
 \bar{\Sigma}^\mu=
 \left(
 \begin{array}{cccc}
  \vec{\varepsilon}\,^{(+)}&\sqrt{2/3}\vec{\varepsilon}\,^{(0)}
   &\sqrt{1/3}\vec{\varepsilon}\,^{(-)} &0 \\
  0 &\sqrt{1/3}\vec{\varepsilon}\,^{(+)} &\sqrt{2/3}\vec{\varepsilon}\,^{(0)} &\vec{\varepsilon}\,^{(-)} \\
 \end{array}
 \right)^\mu ,
\end{align}
and
$\vec{\Sigma}$ is defined by
$\vec{\Sigma}=\frac{3}{2}i\vec{\bar{\Sigma}}\times\vec{\bar{\Sigma}}\,^\dagger$.
The functions $C(r)$ and $T(r)$ are given by
\begin{align}
 &C(r)=\int\frac{d^3 q}{(2\pi)^3}
 \frac{m^2_\pi}{\vec{q}\,^2+m^2_\pi}e^{i\vec{q}\cdot\vec{r}}F(\Lambda, \vec{q}\,),
 \label{eq:central_force} \\
 &S_{\cal O}(\hat{r})T(r)=\int\frac{d^3 q}{(2\pi)^3}
\frac{-\vec{q}\,^2}{\vec{q}\,^2+m^2_\pi}S_{\cal O}(\hat{q})e^{i\vec{q}\cdot\vec{r}}F(\Lambda, \vec{q}\,),
\end{align}
with the form factor \eqref{eq:formfactor}.
We note that the contact term of the central
force~\eqref{eq:central_force} is neglected
as discussed in the nucleon-nucleon meson exchange potential~\cite{Machleidt:1987hj}.

The kinetic terms are give by
\begin{align}
 K_{i}=-\frac{1}{2\mu_i}\triangle_{L_i}+\Delta m_i,
\end{align}
of the channel $i$ given in Table~\ref{table:MBchannels}.
We define the reduced mass $\mu_i=m_{M_i} m_{B_i}/(m_{M_i}+m_{B_i})$ of the meson $M_i(=\bar{D},\bar{D}^\ast)$ and
baryon $B_i(=\Lambda_{\rm c},\Sigma_{\rm c},\Sigma^\ast_{\rm c})$,
$\triangle_{L_i}=\partial^2/\partial r^2+(2/r)\partial/\partial r+L_i(L_i+1)/r^2$
with the orbital angular momentum $L_i$,
and $\Delta m_i=(m_{M_i}+m_{B_i})-(m_{\bar{D}}+m_{\Lambda_{\rm c}})$.

  \section{Computation of spectroscopic factor}
  \label{sec:S-factor}
The wave function of the hidden-charm five-quark ($5q$) state is 
written by three light quarks $uud$ and charm and anti-charm quarks
$c\bar{c}$ as $\Ket{5q}=\Ket{u(1)u(2)d(3)c(4)\bar{c}(5)}$
with the particle number assignment.
The wave function can also be decomposed into various meson-baron components as
\begin{align}
 \Ket{5q}&=a\Ket{\left(u(1)u(2)c(4)\right)^{\frac{1}{2}}\left(d(3)\bar{c}(5)\right)^{0}}+\dots
 \equiv a\Ket{\Sigma^{++}_{\rm c}\bar{D}^{-}}+\dots ,
\end{align}
where $a$ is the definition of the spectroscopic factor~\cite{Hosaka:2004bn}, and
the superscript is the total spin of three quarks or quark-antiquark.
Assuming that
$\Ket{\left(u(1)u(2)c(4)\right)^{\frac{1}{2}}\left(d(3)\bar{c}(5)\right)^{0}}$
is exactly the same as the hadronic wave function of
$\Sigma^{++}_{\rm c}\bar{D}^{-}$,
the spectroscopic factor for the $\Sigma^{++}_{\rm c}\bar{D}^{-}$
channel is
obtained by the overlap
\begin{align}
 a&=\Braket{\Sigma^{++}_{\rm c}\bar{D}^{-}| 5q 
 }.
\label{eq:overlap_appendix}
\end{align}

In this Appendix, we will focus on
the color-flavor-spin wave function of the $5q$ states, 
in which the $uud$ ($3q$) system and the $c\bar{c}$ system are both
in the color octet, and the total color wave function is in the
color-singlet\footnote{The case that the $uud$ system and the $c\bar{c}$ system are both
in the color singlet corresponds to the $J/\psi p$ system.}.
Moreover, the light quarks are assumed to be the $S-$wave state, that is,
the orbital wave function is totally symmetric.
Since the total wave function of the three light quarks must be antisymmetric,
it is represented in Young tableaux as
\begin{align}
 \yng(1,1,1)\,\raisebox{-7mm}{$csfo$} 
  \;\;=\;\;  \yng(1,1,1)\,\raisebox{-7mm}{$csf$} 
 \;\; 
 \cdot
 \;\; 
 \yng(3)\,\raisebox{-3mm}{$o$} 
 \;\;\;\;,
 \label{csfo wf}
\end{align}
where the subscripts $c$, $s$, $f$, and $o$ denote color, spin, flavor,
and orbital wave functions, respectively.
The center dot ``\,$\cdot$\,'' denotes the inner product of wave functions in different
functional space.

The $csf$ wave function is decomposed into color and spin-flavor parts.
In the Young tableaux with the particle number assignment, one obtains (see, e.g., Ref.~\cite{Chen:2002gd})
\begin{align}
 \young(1,2,3)\,\raisebox{-7mm}{$csf$} 
 &= \frac{1}{\sqrt{2}} \left( \;  \young(12,3)\,\raisebox{-5mm}{$c$} 
 \cdot\, 
 \young(13,2)\,\raisebox{-5mm}{$sf$} 
 - \young(13,2)\,\raisebox{-5mm}{$c$} 
 \cdot\,
 \young(12,3)\,\raisebox{-5mm}{$sf$} 
 \; \right).
 \label{3qwf}
\end{align}
In Eq.~\eqref{3qwf}, the color wave functions in the first and second
terms have different types of symmetry for exchanges, 
\begin{equation}
 \young(12,3)\,\raisebox{-5mm}{$c$} 
  \equiv ([21]1)_{c} ,
  \label{color1}
\end{equation}
and
\begin{equation}
 \young(13,2)\,\raisebox{-5mm}{$c$} 
  \equiv ([21]2)_{c} ,
\label{color2}
 \end{equation}
 where $c$  means that  the permutations $[21]1$ and $[21]2$ are 
 performed in the color space.
 The difference between \eqref{color1} and \eqref{color2} lies in the permutation symmetry 
 for exchange: 
 in Eq.~\eqref{color1}, particles 1 and 2 are symmetric for exchange, while particle 1 and 2 are antisymmetric in Eq.~\eqref{color2}.
 The wave function of the $5q$ 
 state is given by the direct product between 
 the $3q$ and $c\bar{c}$ wave functions.
 For this reason, the color part of the total $5q$ state 
 wave function also contains these
 two permutation symmetries, the $([21]1)_{c} $ and the $([21]2)_{c} $, and 
 so in the calculations of the spectroscopic factors,  
 both  permutations  will be considered.

 Since the spin of the $c\bar{c}$ pair can be $S_{c\bar{c}}=0$ or 1,
 there are two 
 $5q$ state 
 wave functions 
 denoted with 
 $\Ket{5q, 
 \bold{S_{c\bar{c}}=0}}$ and
 $\Ket{5q, 
 \bold{S_{c\bar{c}}=1}}$.
 In the case of $S_{c\bar{c}}=0$, the $c\bar{c}$ wave function
 $\psi^{s=0}_{c\bar{c}}$ is
 \begin{align}
  \psi_{c\bar {c}}^{s=0} \sim  \young(54,5)\,\raisebox{-5mm}{$c$}
  \;\; \cdot \;\; 
  \young(4,5)\,\raisebox{-5mm}{$s$} , 
  \label{cc0}
 \end{align}
 and the $5q$ state wave function
 $\Ket{5q, 
 \bold{S_{c\bar{c}}=0}}$ is given by
 \begin{align}
  \Ket{5q, 
  \bold{S_{c\bar{c}}=0}}\; \sim&
  \frac{1}{\sqrt{2}} \left( \; 
  \young(12,3)\,\raisebox{-5mm}{$c$} 
  \;\;\cdot \;\;  
  \young(13,2)\,\raisebox{-5mm}{$sf$} 
  - \young(13,2)\,\raisebox{-5mm}{$c$} 
  \;\;\cdot\;\; 
  \young(12,3)\,\raisebox{-5mm}{$sf$} 
  \; \right) \notag\\
  & \;\;\cdot \;\; 
  \left( \;\;\; 
  \young(54,5)\,\raisebox{-5mm}{$c$} 
  \;\; \cdot \;\; 
  \young(4,5)\,\raisebox{-5mm}{$s$} 
  \;\;\;\right).
  \label{P0}
 \end{align}
 Similarly, the $c\bar{c}$ wave function with spin-triplet,
 $\psi^{s=1}_{c\bar{c}}$, and
 the $5q$ state 
 wave function, $\Ket{5q, 
 \bold{S_{c\bar{c}}=1}}$, are written by
 \begin{align}
  \psi_{c\bar {c}}^{s=1} \sim  \young(54,5)\,\raisebox{-5mm}{$c$}
  \;\;\cdot\;\; 
  \young(45) \,\raisebox{-3mm}{$s$}, 
  \label{cc1}
 \end{align}
 and
 \begin{align}
  \Ket{5q, 
  \bold{S_{c\bar{c}}=1}}\; \sim& 
  \frac{1}{\sqrt{2}} \left( \; 
  \young(12,3)\,\raisebox{-5mm}{$c$} 
  \;\;\cdot\;\; 
  \young(13,2)\,\raisebox{-5mm}{$sf$} 
  - \young(13,2)\,\raisebox{-5mm}{$c$} 
  \;\;\cdot\;\; 
  \young(12,3)\,\raisebox{-5mm}{$sf$} 
  \; \right) \notag\\
  &\otimes \left(\; \;\; 
  \young(54,5)\,\raisebox{-5mm}{$c$} 
  \;\;\cdot\;\; 
  \young(45)\,\raisebox{-3mm}{$s$} 
  \;\;\; \right).
  \label{P1}
 \end{align}

First, let us focus on the term with permutation $([21]1)_{c}$.
 The part of the $5q$ state 
 wave function which contains the permutation
 $([21]1)_{c}$ is 
 \begin{equation}
 \frac{1}{\sqrt{2}} \left( \; 
  \young(12,3)\,\raisebox{-5mm}{$c$} 
  \;\;\cdot\;\; 
  \young(13,2)\,\raisebox{-5mm}{$sf$}
  \; \right) \;\;\cdot\;\;  
 \left( \;\;\; 
  \young(54,5)\,\raisebox{-5mm}{$c$} 
  \;\;\cdot\;\; 
  (S_{c\bar{c}})
  \;\;\;\right),
 \label{P4}
 \end{equation}
 where the $c\bar{c}$ spin part $(S_{c\bar{c}})$ is
 $\young(4,5)\,\raisebox{-5mm}{$s$}$
 or $\young(45)\,\raisebox{-3mm}{$s$}$.
The spin-flavor wave function of the three light quark part in Eq.~\eqref{P4}
can be decomposed into 
\begin{align}
 \yng(2,1)\,\raisebox{-5mm}{$sf$} 
  =\,& \yng(3)\,\raisebox{-3mm}{$f$} 
  \;\;\cdot\;\; 
 \yng(2,1)\,\raisebox{-5mm}{$s$} 
  +\yng(2,1)\,\raisebox{-5mm}{$f$} 
 \;\;\cdot\;\; 
 \yng(3)\,\raisebox{-3mm}{$s$} 
 +\yng(2,1)\,\raisebox{-5mm}{$f$} 
 \;\;\cdot\;\; 
 \yng(2,1)\,\raisebox{-5mm}{$s$} 
 + \yng(1,1,1)\,\raisebox{-7mm}{$f$} 
 \;\;\cdot\;\; 
 \yng(2,1)\,\raisebox{-5mm}{$s$}. 
\label{sfdec}
\end{align}
Assuming that the 
$3q$ 
state belongs to the flavor octet $[21]_{8}$, 
there are two possible 
spin wave functions, 
$[21]_{s}$  and 
$[3]_{s}$,
from Eq.~\eqref{sfdec}. 
In the Young tableaux with particle assignment,
Eq.~\eqref{sfdec} can be expressed as
\begin{align}
 \young(13,2)\,\raisebox{-5mm}{$sf$} 
  = -\frac{1}{\sqrt{2}}\;\left(\young(12,3)\,\raisebox{-5mm}{$f$} 
  \;\;\cdot\;\;
  \young(13,2)\,\raisebox{-5mm}{$s$} 
  +\young(13,2)\,\raisebox{-5mm}{$f$} 
  \;\;\cdot\;\; 
 \young(12,3)\,\raisebox{-5mm}{$s$} 
  \right),
  \label{poss1a}
\end{align}
for the three light quark with spin $\frac{1}{2}$, and
\begin{equation}
    \young(13,2)\,\raisebox{-5mm}{$sf$} 
     = \young(13,2)\,\raisebox{-5mm}{$f$}
     \;\;\cdot\;\; 
     \young(123)\,\raisebox{-3mm}{$s$}, 
  \label{poss2a}
\end{equation}
for the three light quark with spin $\frac{3}{2}$.

Finally, the $5q$ state 
wave function is obtained by combining the 
$3q$ 
and $c\bar{c}$ wave functions.
Since there are different spin configurations for
$3q$ and $c\bar{c}$,
namely $S_{3q}=\frac{1}{2}$ or $\frac{3}{2}$, and $S_{c\bar{c}}=0$ or $1$,
there are several allowed configurations.
\begin{itemize}
 \item[1.] $(S_{c\bar{c}},S_{3q})=(0,\frac{1}{2})$ for $S_{tot}=\frac{1}{2}$ 
\end{itemize}
 By the substitution of Eq.~\eqref{poss1a} into Eq.~\eqref{P4}, we get 
\begin{align}
 \Ket{5q\, ([21]1,1)
 } =& \frac{1}{\sqrt{2}} \left[ \;
 \young(12,3)\,\raisebox{-5mm}{$c$} 
 \;\;\cdot\;\; 
 \left(-\frac{1}{\sqrt{2}}\;\left(
 \young(12,3)\,\raisebox{-5mm}{$f$} 
 \;\;\cdot\;\; 
 \young(13,2)\,\raisebox{-5mm}{$s$} 
 +\young(13,2)\,\raisebox{-5mm}{$f$} 
 \;\;\cdot\;\; 
 \young(12,3)\,\raisebox{-5mm}{$s$} 
 \right) \right) \right]  \notag\\
 &\;\;\cdot\;\; 
 \left( \;\;\; 
 \young(54,5)\,\raisebox{-5mm}{$c$} 
 \;\;\cdot\;\; 
 \young(4,5)\,\raisebox{-5mm}{$s$} 
 \;\;\;\right) \notag\\
 =&-\frac{1}{2} \left[ \;
 \young(12,35,45)\,\raisebox{-7mm}{$c$} 
 \;\;\cdot\;\; 
 \left(
 \young(12,3)\,\raisebox{-5mm}{$f$} 
 \;\;\cdot\;\; 
 \young(13,2)\,\raisebox{-5mm}{$s$} 
 +\young(13,2)\,\raisebox{-5mm}{$f$} 
 \;\;\cdot\;\; 
 \young(12,3)\,\raisebox{-5mm}{$s$} 
 \right) \right] \;\;\cdot\;\; 
 \left( \;\;\;
 \young(4,5)\,\raisebox{-5mm}{$s$} 
 \;\;\;\right).
 \label{f-rearr} 
\end{align}
Herein, $S_{tot}$ is the total spin of the $5q$ state with the quark configuration $(S_{c\bar{c}},S_{3q})$.
We also introduce the notation 
$\Ket{5q\,({[21]m,n})}$ 
to identify the $5q$ state wave function which comes from the color part
$m=1,2$ while  $n=1,2,3,4$ is the index of the channels, $(S_{c\bar{c}},S_{3q})=(0,\frac{1}{2})$,
$(0,\frac{3}{2})$, $(1,\frac{1}{2})$ and $(1,\frac{3}{2})$, respectively.
\begin{itemize}
 \item[2.] $(S_{c\bar{c}},S_{3q})=(1,\frac{1}{2})$ for $S_{tot}=\frac{1}{2}$ or $\frac{3}{2}$
\end{itemize}
In a similar to Eq.~\eqref{f-rearr}, we get 
\begin{align}
\Ket{5q\,({c,[21]1,2})} 
 =&-\frac{1}{2} \left[ \;
 \young(12,35,45)\,\raisebox{-7mm}{$c$} 
 \;\;\cdot\;\; 
 \left(
 \young(12,3) \,\raisebox{-5mm}{$f$} 
 \;\;\cdot\;\; 
 \young(13,2)\,\raisebox{-5mm}{$s$} 
 +\young(13,2)\,\raisebox{-5mm}{$f$} 
 \;\;\cdot\;\; 
 \young(12,3)\,\raisebox{-5mm}{$s$} 
 \right) \right] \;\;\cdot\;\; 
 \left( \;\;\;  \young(45)\,\raisebox{-3mm}{$s$} 
 \;\;\;\right).
 \label{f-rearr2}
\end{align}
\begin{itemize}
 \item[3.] $(S_{c\bar{c}},S_{3q})=(0,\frac{3}{2})$ for $S_{tot}=\frac{3}{2}$
\end{itemize}
By the substitution of Eq.~\eqref{poss2a} into Eq.~\eqref{P4}, we get 
\begin{align}
 \Ket{5q\,({[21]1,3})} 
 =&\frac{1}{\sqrt{2}} \left[ \; 
 \young(12,35,45)\,\raisebox{-7mm}{$c$} 
 \;\;\cdot\;\; 
 \left(
 \young(13,2)\,\raisebox{-5mm}{$f$} 
 \;\;\cdot\;\; 
 \young(123)\,\raisebox{-3mm}{$s$} 
 \right) \right] \;\;\cdot\;\; 
 \left( \;\;\; 
  \young(4,5)\,\raisebox{-5mm}{$s$} 
 \;\;\;\right) .
   \label{f-rearr3}
\end{align}
\begin{itemize}
 \item[4.] $(S_{c\bar{c}},S_{3q})=(1,\frac{3}{2})$ for $S_{tot}=\frac{1}{2}$, $\frac{3}{2}$, or $\frac{5}{2}$
\end{itemize}
In a similar way to Eq.~\eqref{f-rearr3},
we get 
\begin{align}
 \Ket{5q\,({[21]1,4})} 
 =&\frac{1}{\sqrt{2}} \left[ \; 
 \young(12,35,45) \,\raisebox{-7mm}{$c$} 
 \;\;\cdot\;\; 
 \left(
 \young(13,2)\,\raisebox{-5mm}{$f$} 
 \;\;\cdot\;\; 
 \young(123) \,\raisebox{-3mm}{$s$} 
 \right) \right] 
 \;\;\cdot\;\; 
 \left( \;\;\; 
  \young(45)\,\raisebox{-3mm}{$s$} 
 \;\;\;\right) .
 \label{f-rearr4}
\end{align}

The spin part needs one more step. 
For instance, in the case number 3 
for $\Ket{5q\,({[21]1,3})}$,
the spin wave function has the coupling structure with $S_{123}=S_{3q}=\frac{3}{2}$ and
$S_{45}=S_{c\bar{c}}=0$ as
\begin{equation}
[(S_{13}\otimes S_{2})^{S_{123}}\otimes(S_{4}\otimes S_{5})^{S_{45}} ]^{S_{tot}}= 
\left[\left(1\otimes\frac{1}{2}\right)^{\frac{3}{2}}\otimes \left(\frac{1}{2}\otimes \frac{1}{2}
\right)^{0}
\right]^{\frac{3}{2}},
\end{equation}
which 
is recoupled for  the channel of the $\Sigma^{(\ast)}_{c}$ baryon  and the $\bar{D}^{(\ast)}$ meson 
by the spin rearrangement 
\begin{equation}
 \left[\left(1\otimes\frac{1}{2}\right)^{\frac{3}{2}}\otimes \left(\frac{1}{2}\otimes \frac{1}{2}
\right)^{0}
\right]^{\frac{3}{2}} = 
\sum_{s_{134},s_{25}}C_{s_{134},s_{25}}
\left[\left(1\otimes\frac{1}{2}\right)^{S_{134}}\otimes \left(\frac{1}{2}\otimes \frac{1}{2}
\right)^{S_{25}}
\right]^{\frac{3}{2}},
\end{equation}
where 
\begin{equation}
C_{\frac{1}{2},1}= -\frac{1}{\sqrt{3}}, \;\;\;\;\;C_{\frac{3}{2},0}=\frac{1}{2},   
\;\;\;\;\; C_{\frac{3}{2},1}=\frac{1}{2} \sqrt{\frac{5}{3}}  .
\end{equation}
Here, the coefficients $C_{\frac{1}{2},1}$, $C_{\frac{2}{2},0}$ and $C_{\frac{3}{2},1}$ 
are the amplitude for the  spin  components 
$(S_{134},S_{25})=(\frac{1}{2},1)$,
$
(\frac{3}{2},0)$, 
and $
(\frac{3}{2},1)$, respectively, 
which 
correspond to the $\Sigma_{c}\bar{D}^{*}$, $\Sigma^{*}_{c}\bar{D}$, and
$\Sigma_{c}^{*}\bar{D}^{*}$ baryon-meson channel, respectively. 
From Eq.~\eqref{f-rearr3},
one finds the
amplitude of 
the each baryon-meson components in 
$\Ket{5q\,({[21]1,3})}$,
\begin{equation}
\Ket{5q\,({[21]1,3})} 
=-\frac{1}{\sqrt{6}}\Ket{\Sigma_{c}\bar{D}^{*}} 
+\frac{1}{2\sqrt{2}}\Ket{\Sigma_{c}^{*}\bar{D}}
+\frac{1}{2}\sqrt{\frac{5}{6}}\Ket{\Sigma_{c}^{*}\bar{D}^{*}}+...
\label{eq:s-factor_211_3}
\end{equation}
From Eqs.~\eqref{eq:overlap_appendix} and~\eqref{eq:s-factor_211_3},
the spectroscopic factor is obtained.

In a way similar to the permutation $([21]1)_{c}$, the wave function for $([21]2)_{c}$ can be obtained.
 The part of the $5q$ state wave function which contains the permutation $([21]2)_{c}$ is 
 \begin{equation}
 -\frac{1}{\sqrt{2}} \left( \; 
  \young(13,2) \,\raisebox{-5mm}{$c$} 
  \;\;\cdot\;\; 
  \young(12,3)\,\raisebox{-5mm}{$sf$} 
  \; \right) \;\;\cdot\;\; 
 \left( \;\;\; 
  \young(54,5)\,\raisebox{-5mm}{$c$} 
  \;\;\cdot\;\;  
  \young(4,5)\,\raisebox{-5mm}{$s$} 
  \;\;\;\right),
 \label{P4b}
\end{equation}
for the $c\bar{c}$ pair in the singlet state and
\begin{equation}
  -\frac{1}{\sqrt{2}} \left( \; 
   \young(13,2)\,\raisebox{-5mm}{$c$} 
   \;\;\cdot\;\; 
   \young(12,3)\,\raisebox{-5mm}{$sf$} 
   \; \right)
  \;\;\cdot\;\; 
  \left(\; \;\; 
  \young(54,5)\,\raisebox{-5mm}{$c$} 
  \;\;\cdot\;\; 
  \young(45)\,\raisebox{-3mm}{$s$} 
  \;\;\; \right),
  \label{P5b}
\end{equation}
for the $c\bar{c}$ pair in the triplet state.
In the Young tableaux with particle assignment, 
the spin-flavor decomposition of Eq.~\eqref{sfdec} 
can be expressed as 
\begin{equation}
  \young(12,3)\,\raisebox{-5mm}{$sf$} 
   = \frac{1}{\sqrt{2}}\;\left(\young(12,3)\,\raisebox{-5mm}{$f$} 
  \;\;\cdot\;\; 
  \young(12,3)\,\raisebox{-5mm}{$s$} 
  -\young(13,2)\,\raisebox{-5mm}{$f$} 
  \;\;\cdot\;\; 
  \young(13,2)\,\raisebox{-5mm}{$s$} 
			 \right),
  \label{poss1ab}
\end{equation}
for the three light quark with spin $\frac{1}{2}$ and
\begin{equation}
  \young(12,3)\,\raisebox{-5mm}{$sf$} 
   = \young(12,3)\,\raisebox{-5mm}{$f$} 
  \;\;\cdot\;\; 
  \young(123)\,\raisebox{-3mm}{$s$} , 
  \label{poss2ab}
\end{equation}
for the three light quark with spin $\frac{3}{2}$.
As in the case of the color permutation $[21]1$,
from the combination of the $3q$ 
and 
$c\bar{c}$ 
wave functions, 
several allowed 
configurations have to be considered. 
\begin{itemize}
 \item[1.] $(S_{c\bar{c}},S_{3q})=(0,\frac{1}{2})$ for $S_{tot}=\frac{1}{2}$
\end{itemize}
By the substitution of Eq.~\eqref{poss1ab} into Eq.~\eqref{P4b} we get 
 \begin{equation}
\Ket{5q\,({[21]2,1})}
   =-\frac{1}{2} \left[ \; 
		  \young(13,25,45)\,\raisebox{-7mm}{$c$} 
		  \;\;\cdot\;\; 
		  \left(\young(12,3)\,\raisebox{-5mm}{$f$} 
		   \;\;\cdot\;\; 
		   \young(12,3)\,\raisebox{-5mm}{$s$} 
		   -\young(13,2)\,\raisebox{-5mm}{$f$} 
		   \;\;\cdot\;\; 
		   \young(13,2)\,\raisebox{-5mm}{$s$} 
		  \right) \right] 
   \;\;\cdot\;\; 
   \left( \;\;\; 
   \young(4,5)\,\raisebox{-5mm}{$s$} 
   \;\;\;\right) .
   \label{f-rearrb}
  \end{equation}
\begin{itemize}
 \item[2.] $(S_{c\bar{c}},S_{3q})=(1,\frac{1}{2})$ for $S_{tot}=\frac{1}{2}$ or $S_{tot}=\frac{3}{2}$
\end{itemize}
By the substitution of Eq.~\eqref{poss1ab} into Eq.~\eqref{P5b} we get 
 \begin{equation}
  \Ket{5q\,({[21]2,2})}
   =-\frac{1}{2} \left[ \; 
		  \young(13,25,45)\,\raisebox{-7mm}{$c$} 
		  \;\;\cdot\;\; 
		  \left(\young(12,3)\,\raisebox{-5mm}{$f$} 
		   \;\;\cdot\;\;  
		   \young(12,3)\,\raisebox{-5mm}{$s$} 
		   -\young(13,2)\,\raisebox{-5mm}{$f$} 
		   \;\;\cdot\;\;     
		   \young(13,2)\,\raisebox{-5mm}{$s$} 
		  \right) \right] 
   \;\;\cdot\;\; 
   \left( \;\;\; 
   \young(45)\,\raisebox{-3mm}{$s$} 
   \;\;\;\right) .
   \label{f-rearr2b}
  \end{equation}
\begin{itemize}
   \item[3.] $(S_{c\bar{c}},S_{3q})=(0,\frac{3}{2})$ for $S_{tot}=\frac{3}{2}$
\end{itemize}
 By the substitution of Eq.~\eqref{poss2ab} into Eq.~\eqref{P4b} we get 
 \begin{equation}
\Ket{5q\,({[21]2,3})}
   =-\frac{1}{\sqrt{2}} \left[ \; 
			 \young(13,25,45)\,\raisebox{-7mm}{$c$} 
			 \;\;\cdot\;\;  
			 \left(\young(12,3)\,\raisebox{-5mm}{$f$} 
			  \;\;\cdot\;\; 
			  \young(123)\,\raisebox{-3mm}{$s$} 
			 \right) \right] 
   \;\;\cdot\;\; 
   \left( \;\;\; 
    \young(4,5)\,\raisebox{-5mm}{$s$} 
   \;\;\;\right) .
   \label{f-rearr3b}
 \end{equation}
\begin{itemize}
 \item[4.] $(S_{c\bar{c}},S_{3q})=(1,\frac{3}{2})$ for $S_{tot}=\frac{1}{2}$, $\frac{3}{2}$, or $\frac{5}{2}$
\end{itemize}
By the substitution of Eq.~\eqref{poss2ab} into Eq.~\eqref{P5b} we get 
  \begin{equation}
   \Ket{5q\,({[21]2,4})} 
    =-\frac{1}{\sqrt{2}} \left[ \; 
			  \young(13,25,45)\,\raisebox{-7mm}{$c$} 
			  \;\;\cdot\;\;  
			  \left(
			   \young(13,2)\,\raisebox{-5mm}{$f$} 
			   \;\;\cdot\;\; 
			   \young(123)\,\raisebox{-3mm}{$s$} 
			  \right) \right] 
    \;\;\cdot\;\; 
    \left( \;\;\; 
     \young(45)\,\raisebox{-3mm}{$s$} 
     \;\;\;\right) .
    \label{f-rearr4b}
 \end{equation}


 

\end{document}